\documentclass{nature}
\usepackage{caption}
\usepackage{hyperref}
\usepackage{epsfig}
\usepackage{color}
\usepackage{ulem}
\def\mearth{{\rm\,M_\oplus}}
\def\rearth{{\rm\,R_\oplus}}

\newcommand\icarus{{Icarus}}%
\newcommand\mnras{{MNRAS}}%
\newcommand\aap{{A\&A}}%
\newcommand\apj{{ApJ}}%
\newcommand\apjl{{ApJ}}%
\newcommand\nat{{Nature}}%
\newcommand\aj{{AJ}}%
\newcommand\araa{{ARA\&A}}
\newcommand\ssr{{Space~Sci.~Rev.}}
\newcommand\gca{{Geochem.~Cosmochem.~Acta}}%

\title{An upper limit on late accretion and water delivery in the Trappist-1 exoplanet system}

\author{Sean~N.~Raymond$^{1}$, Andre Izidoro$^2$, Emeline Bolmont$^3$, Caroline Dorn$^{4}$, Franck Selsis$^1$, Martin Turbet$^3$, Eric Agol$^5$, Patrick Barth$^{6,7,8}$, Ludmila Carone$^9$, Rajdeep Dasgupta$^2$, Michael Gillon$^{10}$, Simon L. Grimm$^{11}$}

\begin{document}

\maketitle

\begin{affiliations}
 \item Laboratoire d'astrophysique de Bordeaux, Univ. Bordeaux, CNRS, B18N, all{\'e}e Geoffroy Saint-Hilaire, 33615 Pessac, France
\item Department of Earth, Environmental and Planetary Sciences, Rice University, 6100 Main Street, MS 126, Houston, TX 77005, USA
\item Observatoire de Genève, Université de Genève, Chemin Pegasi 51, 1290, Sauverny, Switzerland
\item University of Zurich, Institute of Computational Sciences, Winterthurerstrasse 190, CH-8057, Zurich, Switzerland
\item Department of Astronomy, University of Washington, Seattle, WA 98195-1580, USA
\item Centre for Exoplanet Science, University of St Andrews, North Haugh, St Andrews, KY16 9SS, UK
\item SUPA, School of Physics \& Astronomy, University of St Andrews, North Haugh, St Andrews, KY16 9SS, UK
\item School of Earth \& Environmental Sciences, University of St Andrews, Irvine Building, St Andrews, KY16 9AL, UK
\item Max Planck Institute for Astronomy, K\"onigstuhl 17, 69117 Heidelberg, Germany
\item Space sciences, Technologies and Astrophysics Research (STAR) Institute, Universite de Liege, 4031 Liege, Belgium
\item Center for Space and Habitability, University of Bern, Gesellschaftsstrasse 6, 3012 Bern, Switzerland
\end{affiliations}

\begin{abstract}
The Trappist-1 system contains seven roughly Earth-sized planets locked in a multi-resonant orbital configuration,\cite{gillon17,luger17} which has enabled precise measurements of the planets' masses and constrained their compositions.\cite{agol21} Here we use the system's fragile orbital structure to place robust upper limits on the planets' bombardment histories.  We use N-body simulations to show how perturbations from additional objects can break the multi-resonant configuration by either triggering dynamical instability or simply removing the planets from resonance. The planets cannot have interacted with more than $\mathbf{\sim 5\%}$ of an Earth mass ($\mathbf{M_\oplus}$) in planetesimals -- or a single rogue planet more massive than Earth's Moon -- without disrupting their resonant orbital structure. This implies an upper limit of $\mathbf{10^{-4}}$ to $\mathbf{10^{-2} M_\oplus}$ of late accretion on each planet since the dispersal of the system's gaseous disk.  This is comparable to or less than the late accretion on Earth after the Moon-forming impact\cite{day07,walker09}, and demonstrates that the Trappist-1 planets' growth was complete in just a few million years, roughly an order of magnitude faster than Earth's.\cite{wood05,kleine09} Our results imply that any large water reservoirs on the Trappist-1 planets must have been incorporated during their formation in the gaseous disk. 
\end{abstract}

The seven-planet Trappist-1 system is characterized by mean motion resonances between neighboring planets and 3-body (Laplace) resonances between triplets of adjacent planets (see Methods sections 1 and 3 and Extended Data Fig.~1).\cite{gillon17,luger17} Here we use the multi-resonant structure of the Trappist-1 system to constrain the planets' bombardment histories after their formation. We use N-body simulations to determine the perturbations from additional objects in the system required to break the Trappist-1 resonant chain. Given the survival of the resonant chain from its formation to the present day, we place an upper limit on the integrated mass with which the system has interacted and the impacts that may have taken place. 

Starting from a configuration of the system that is consistent with current constraints and stable for Gyr timescales, we invoke gravitational perturbations from additional objects under two limiting assumptions.  First, we assume that a single rogue planetary embryo from the outer disk was scattered inward toward the planets. 
The planets undergo stochastic exchanges of orbital energy with the rogue planet during close gravitational encounters, which can excite the planets' orbits, disrupt the system's multi-resonant structure and induce instability. 
We ran 1000 simulations varying the mass of a rogue planet in the range $0.001 - 1 \mearth$. Second, we assume that a swarm of leftover planetesimals from the outer disk was scattered inward toward the planets. 
The planets undergo an exchange of orbital angular momentum with the planetesimals, which causes their orbits to spread out.\cite{fernandez84} 
We ran 140 simulations varying the total mass in 1000 equal-mass planetesimals between $0.01 \mearth$ and $3 \mearth$. Planetesimals either represent leftovers whose orbits were excited by the planets' migration\cite{izidoro21} or objects scattered inward toward the known planets, perhaps by a more distant planet.\cite{kral18,dosovic20} While other potential sources of impacts exist\cite{kral18}, we focused on inward-scattered rogue objects because they provide analogs to late accretion on Earth\cite{day07,walker09} and have the highest impact probabilities\cite{kral18} (see Methods Section 4 and Extended Data Table 2).  Each system was integrated using the hybrid {\tt Mercury} integrator\cite{chambers99} for 10 million years or until it became unstable (see Methods Section 2, Extended Data Table 1, and Extended Data Figs.~2 and 3). This is 2-3 orders of magnitude shorter than the lifetime of the system,\cite{gonzales19} meaning that our upper limits are conservative, as some systems may go unstable or break their resonances on longer timescales.  We performed additional calculations to verify that our results are robust to the initial semimajor axis and periastron distribution of rogue objects and to our choice of the ``best-fit'' configurations of the system (see Methods Sections 4 and 5, Extended Data Tables 1 and 2, Supplementary Table 1, and Extended Data Figs.~4 and 5).\cite{agol21} We also calculated that star-planet tidal effects did not shift the outer planets' orbits enough to affect our results (see Methods Section 7 and Extended Data Fig. 7).

The Trappist-1 system is dynamically fragile. While the planets maintained their multi-resonant structure in most simulations containing a rogue planet less massive than $\sim 0.005 \mearth$, more massive rogue planets were disruptive (Fig.~1). The majority of systems with rogue planet masses between $0.005$ and $0.02 \mearth$ disrupted their resonances but remained stable for the 10 million year duration of the simulation.  More massive rogue planets generally triggered dynamical instability, leading to collisions between planets. The simulation with the most massive rogue planet that maintained its stable resonant structure was just over a lunar mass ($\sim 0.01 \mearth$), making this the most massive possible giant impact that could have taken place on any of the Trappist-1 planets at any time since the dispersal of the star's gaseous disk. The three outer planets were most likely to undergo a giant impact, with relative probabilities of 11\% (planet f), 28\% (planet g), and 31\% (planet h).  If such an impact occurred, it would have been the only late giant impact on any of the planets.

\begin{figure*}
  \begin{center} 
 \leavevmode \epsfxsize=8.2cm\epsfbox{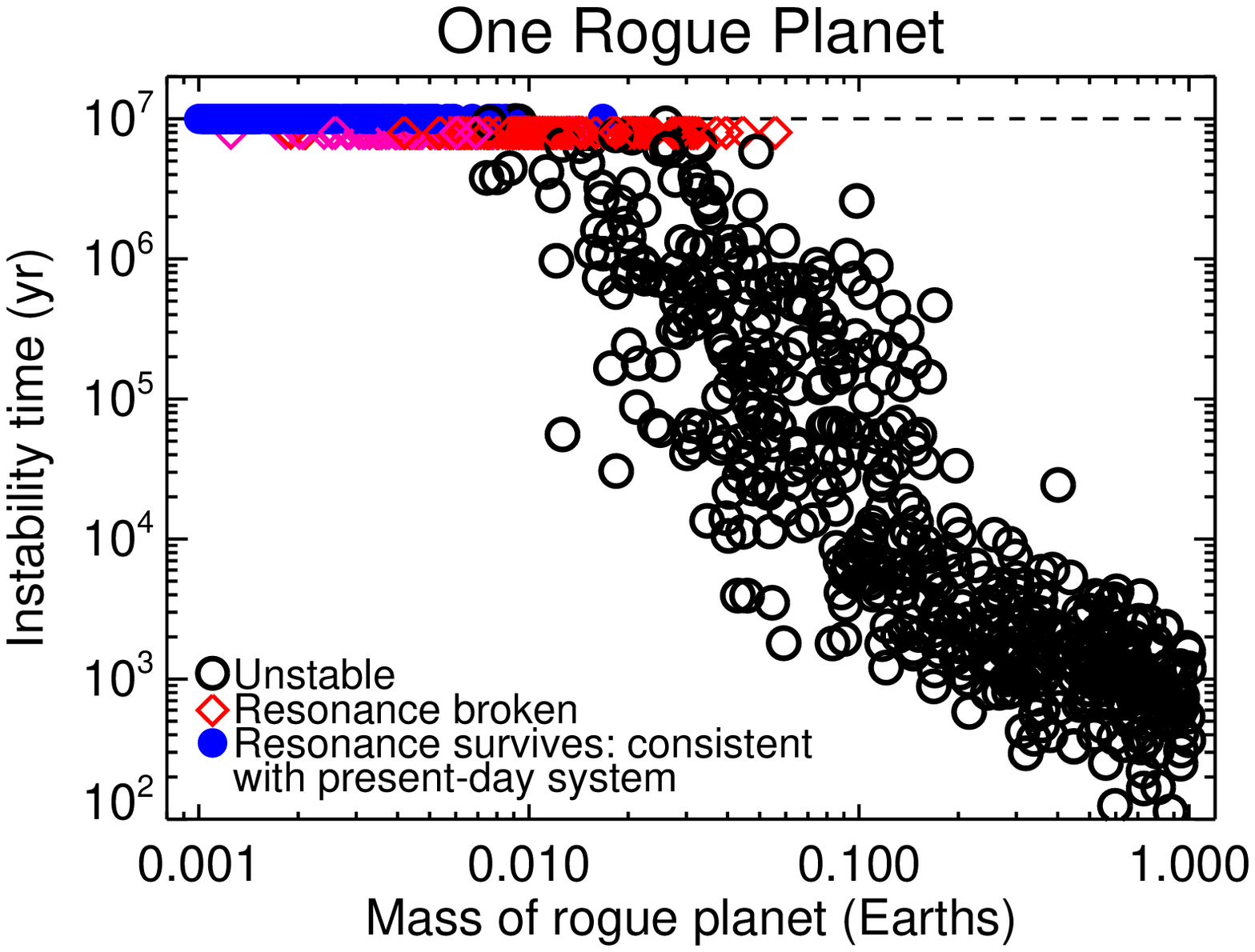}
    \leavevmode \epsfxsize=8.2cm\epsfbox{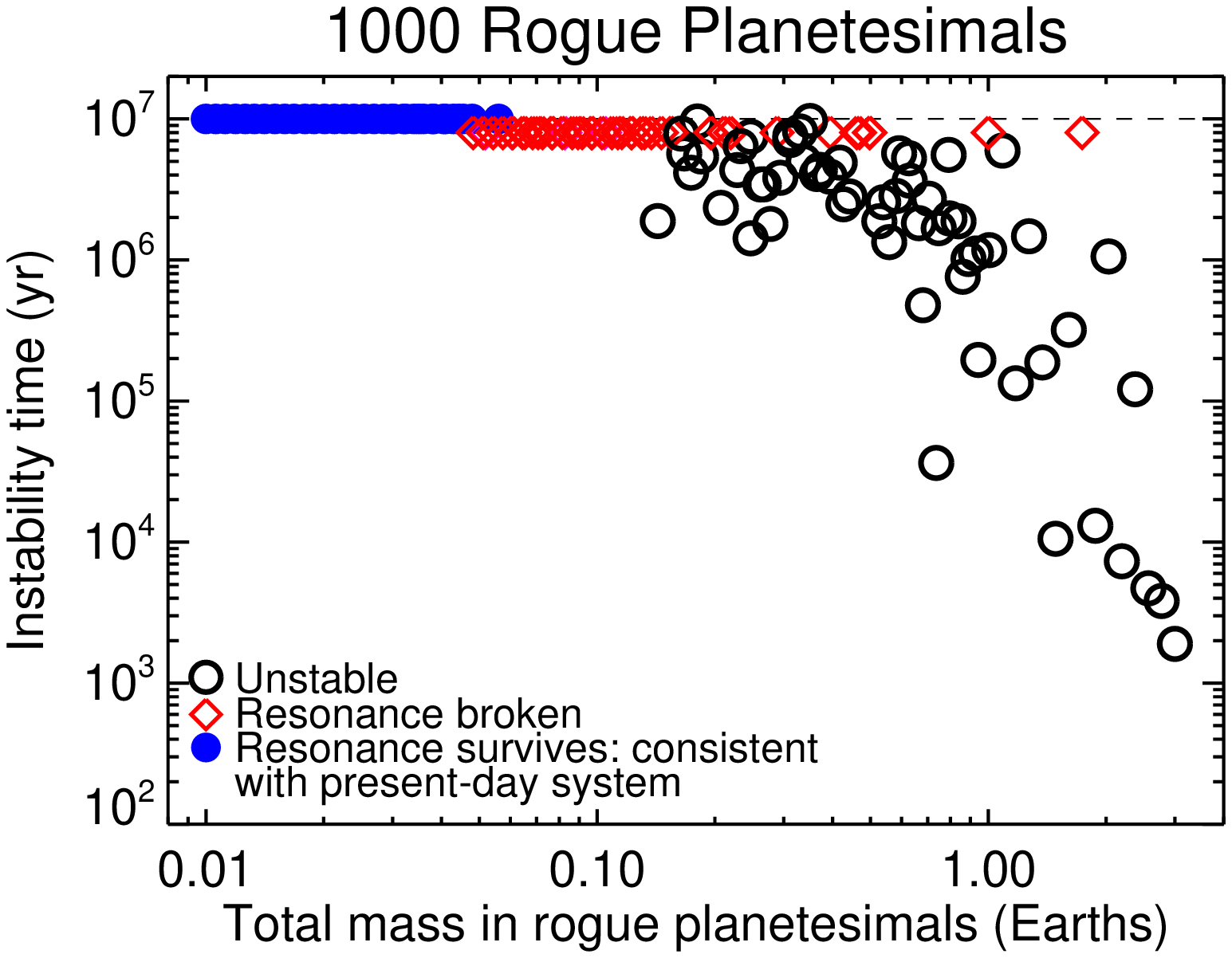}
    \caption[]{{\bf Response of the Trappist-1 system when interacting with a single rogue planet (left panel) or 1000 rogue planetesimals (right panel).} Blue circles represent simulations that are consistent with the observed system: the planets' orbits remained stable for the 10 Myr integration and retained their multi-resonant configuration. The red symbols denote systems that were stable for 10 Myr (shifted slightly downward for visibility) but whose resonant structure was disrupted, as quantified using Laplace resonant angles for the outer triplets of planets (see Methods).  The black circles represent systems that underwent dynamical instabilities, leading to planet-planet collisions. The starting orbits of the rogue planets or planetesimals were chosen assuming that they were scattered inward from just exterior to the known planets. The results of simulations with different configurations of the Trappist-1 planets produced near-identical results (see Methods Section 5).
        } 
     \label{fig:tinst}
    \end{center}
\end{figure*}

A population of rogue planetesimals also easily disrupts the Trappist-1 system (Fig.~1).  The outer planets are the first to interact with the planetesimals, which are for the most part scattered inward. The planets' orbits widen and spread out due to the back-reaction of planetesimal scattering.\cite{fernandez84}  When the planets scatter a total of more than $\sim 0.05 \mearth$ in planetesimals the system's multi-resonant structure is disrupted, as the angles that characterize mean-motion and 3-body resonances transition from libration to circulation.  We used the evolution of $\phi_5$ -- the angle characterizing the observed Laplace resonance between planets f, g, and h\cite{luger17} -- to determine which systems were still in resonance (see Methods Section 3).  After leaving their stabilized resonant state,\cite{tamayo17} most systems that interacted with more than $\sim 0.1 \mearth$ in planetesimals became dynamically unstable and underwent planet-planet collisions. The instability timescale was shorter in systems with more massive swarms of rogue planetesimals, and it is likely that many of the systems from Fig.~1 that left resonance but survived would become dynamically unstable on timescales longer than our 10 Myr integrations. 

The maximum integrated bombardment fluxes on the planets are measured in thousandths of an Earth mass (Table 1). These were calculated by imposing a total planetesimal mass of $0.05 \mearth$ -- the maximum mass for which the resonant structure was preserved (Fig.~1) -- and measuring the relative impact rates on the planets in simulations that maintained their resonant structure. The spreading of the planets' orbits resulted from the cumulative effect of planetesimal perturbations such that the impacts could have taken place at any time during the system's evolution. Given the short collisional timescales of close-in planetesimals,\cite{kral18} we expect most collisions to happen shortly after the dissipation of the gaseous disk. The impact speeds in our simulations were consistent with previous studies for our chosen impactor population,\cite{kral18,dosovic20} with median velocities of 30 km/s for planet b decreasing to 13 km/s for planet h (see Methods Section 6 and Extended Data Fig.~6).


\begin{table*} 
\centering
\begin{tabular}{ c | c | c | c}
Planet & Orbital radius & Maximum bombardment & Maximum water delivered \\
 & (AU) & mass ($\mearth$) & (Earth oceans) \\
  \hline			
  b & 0.0115 & 0.00038 & 0.15\\
  c & 0.0158 & 0.0015 & 0.64 \\
  d & 0.0223 & 0.0016 & 0.68 \\
  e & 0.0293 & 0.0035 & 1.54 \\
  f & 0.0385 & 0.008 & 3.41 \\
  g & 0.0469 & 0.018 & 7.62 \\
  h & 0.0620 & 0.012 & 6.16 \\
  \hline  
\end{tabular}
\label{tab:results}
\caption{Upper limits on the total mass in planetesimals that collided with each of the Trappist-1 planets and the maximum amount of water delivered by those impacts (assuming perfect mergers).  Here we have adopted a maximum value of $0.05 \mearth$ in scattered rogue planetesimals (see Fig.~1) and assumed that they contain 10\% water by mass, typical of the wettest carbonaceous chondrite meteorites.\cite{alexander18}  One ``ocean'' of water is equivalent to the mass of the water on Earth's surface, $2.3 \times 10^{-4} \mearth$. }
\end{table*}

Our results demonstrate that the Trappist-1 planets grew an order of magnitude faster than Earth (Fig.~2). Cosmochemical constraints\cite{wood05,kleine09} and accretion models\cite{morby12b} indicate that a large portion of Earth's growth took place after the dissipation of the Sun's gaseous disk. Mars' growth was close to complete by the time the disk dissipated\cite{dauphas11}, and Earth's constituent embryos were likely $\sim$Mars-sized at that time.\cite{morby12b} Earth's final, Moon-forming impact took place $\sim 50-100$ million years later.\cite{kleine09} The abundance of highly siderophile elements (HSEs) in the mantle and crust suggest that Earth accreted $\sim 0.003-0.007 \mearth$ in planetesimals in the 4.45 Gyr between the Moon-forming impact and the present day;\cite{day07,walker09} this is generally called `late accretion.' Our dynamically-derived upper limits on late accretion on the Trappist-1 planets are more stringent than those for the Earth because the time window extends all the way back to the end of the gaseous disk phase. Nonetheless, we constrain late accretion on the three innermost Trappist-1 planets to have been significantly lower than on Earth (Table~1).  The four outer Trappist-1 planets could not have accreted more than $\sim 1\%$ of their masses during this interval, comparable to upper limits on Earth's late accretion.\cite{marchi18} The Trappist-1 planets must therefore have been fully assembled during the few million year-long gaseous disk phase. This is in line with successful formation models, which invoke gas-driven orbital migration.\cite{ormel17,papaloizou18,coleman19} Migration appears to be essential in producing an orbital configuration with a long dynamical lifetime.\cite{tamayo17} The starting orbital distance of migrating planets is unconstrained, although models expect large embryos to form fastest just beyond the snow line.\cite{ormel17,miguel20} While the low occurrence rate of multi-resonant systems suggest that most resonant chains undergo dynamical instabilities after the dispersal of the disk,\cite{terquem07,izidoro21} Trappist-1 represents a rare surviving resonant chain.


\begin{figure*}
  \begin{center} 
   \leavevmode \epsfxsize=15cm\epsfbox{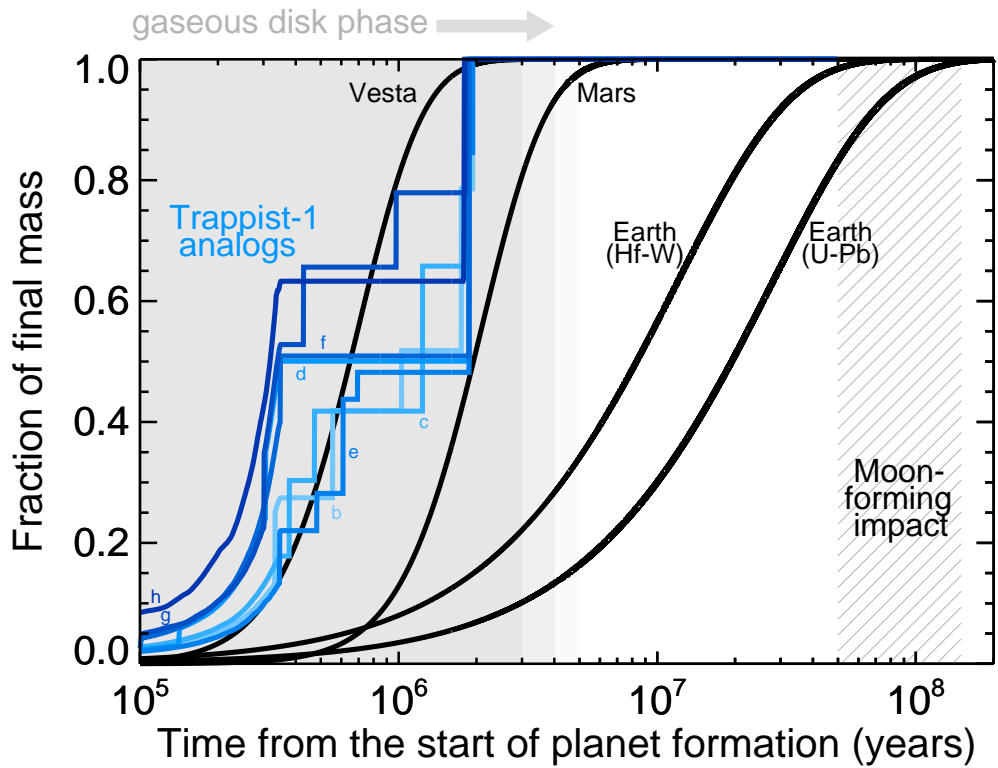}
    \caption[]{{\bf A possible growth history for the Trappist-1 planets compared with constraints for Solar System objects.} Characteristic growth timescales for Vesta, Mars and Earth (and for the Moon-forming impact) come from analyses of different radioactive systems.\cite{wood05,kleine09,dauphas11} After the dispersal of the gaseous disk, Earth's growth was likely characterized by giant impacts\cite{morby12b} (including the Moon-forming impact), which are not reflected in the Hf-W and U-Pb model curves.\cite{wood05} We used a simulation by Izidoro et al\cite{izidoro21} that formed a close-in, long-term stable multi-resonant chain of planets. The growth curves of seven planets with masses between 0.4 and 1.5 $\mearth$ -- characterized by an early phase of pebble accretion followed by giant impacts -- are included as analogs for the accretion of the Trappist-1 planets (see Methods Section 9). Each planet's growth curve is shown in a shade of blue, with darker colors representing more distant orbits. While this is only one illustrative example among many,\cite{coleman19,miguel20} the Trappist-1 system's growth must have completed by the end of the gaseous disk phase, whose duration is constrained by observations of the occurrence rate of hot dust around stars in clusters of different ages.\cite{mamajek09} 
    }
     \label{fig:growth}
    \end{center}
\end{figure*}

To investigate the role of late water delivery to the Trappist-1 planets, we performed new calculations to estimate the planets' water contents. We used interior structure\cite{Dorn17} and atmospheric\cite{turbet20aa} models constrained by the latest planetary mass and radius estimates.\cite{agol21} Given that inferred water contents are subject to interior degeneracies,\cite{dorn18,agol21,unterborn18,barth20,acuna21} we estimated water mass fractions assuming five different rocky interior models (see Methods). Across all of our models we found that planets b, c and d are likely to be volatile-depleted whereas the outer planets (e through h) are more likely to have significant water contents (consistent with previous models\cite{acuna21,agol21}). Yet while our mean water contents are $\sim$1-10\% wt\% for the outer planets, 2-$\sigma$ error bars reach down to zero in at least one model for each planet (Fig.~3). We therefore cannot be certain of a water-rich composition for any of the planets. Planet g is our strongest candidate for being water-rich, with just one model reaching zero water content at the $2 \sigma$ level. 


\begin{figure*}
  \begin{center} 
   \leavevmode \epsfxsize=15cm\epsfbox{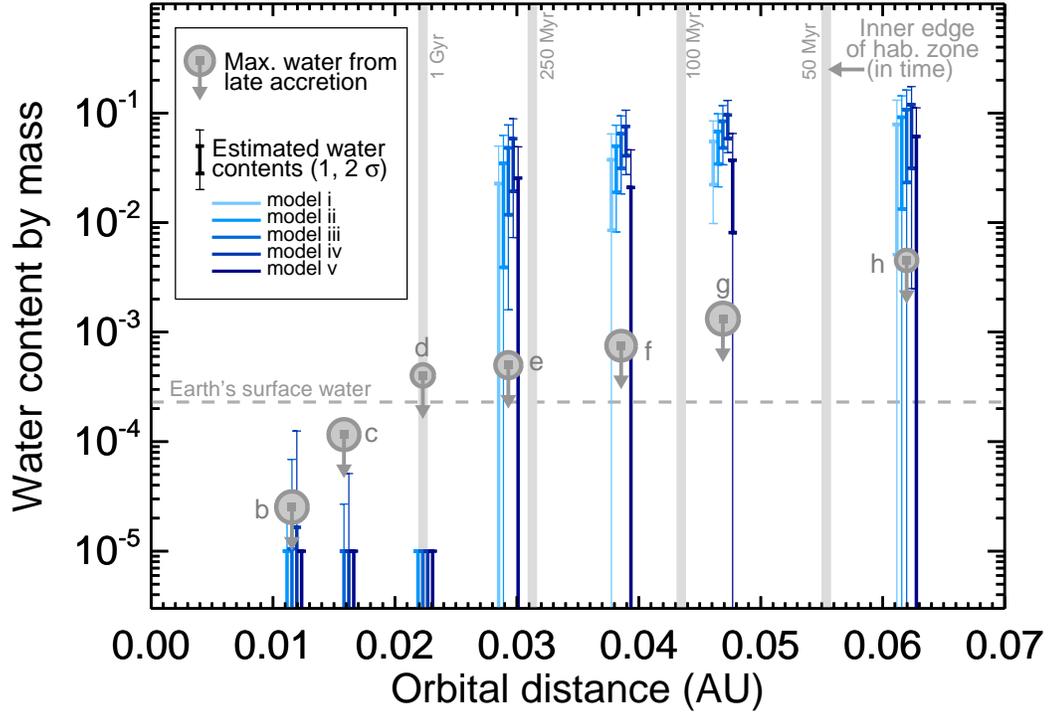}
    \caption[]{{\bf Late water delivery to the Trappist-1 planets.} The thick and thin error bars indicate the one and two-sigma confidence intervals (encompassing 68\% and 95\% probability), respectively, for estimates of the bulk water content of each planet, for each of the five models discussed in the text (note that certain models do not apply to the planets b, c and d; see Methods Section 8). Each circle is proportional to the planet's size, and its vertical position represents an upper limit on the water delivered during late accretion. The grey vertical lines show snapshots of the location of the inner edge of the habitable zone calculated with 3D simulations,\cite{turbet18} which swept inward in time as the central star evolved to the main sequence.\cite{ramirez14,tian15} Earth's surface water budget of one ``ocean'' is shown as the dashed horizontal line. 
        }
     \label{fig:water}
    \end{center}
\end{figure*}

Late accretion may have delivered the full present-day water budgets of planets b, c, and d (Fig.~3). These upper limits assume that late impactors contained 10\% water by mass, typical of the wettest carbonaceous chondrite meteorites.\cite{alexander18} They also assume perfect merging and no water loss, although only a fraction of material is likely to be accreted (as little as 1-10\% for the highest-velocity impacts).\cite{kral18,schlichting15} Taking the upper limits at face value, late accretion can deliver enough water for planets e through h to each have bulk water contents similar to Earth's value of $\sim 10^{-3}$ by mass.\cite{marty12} However, late accretion falls an order of magnitude short of delivering enough water to explain the mean water contents from our models (Fig.~3). If future work confirms water contents at the wt\% level or higher for these planets, then a separate source of water would be required beyond late accretion. The most natural source is water incorporated during the planets' accretion, as predicted by formation models.\cite{miguel20,izidoro21,ormel17,unterborn18,coleman19} To acquire significant water on planets e-h requires long-range orbital migration.  Some constituent pieces of each planet must have originated beyond the snow line, which was more distant during the star's brighter early phases.\cite{mulders15c,unterborn18} Two viable possibilities are (i) that the cores of the planets originated just past the snow line and migrated inward\cite{ormel17}, or (ii) that the snow line swept inward, allowing rocky cores to efficiently accrete icy pebbles and migrate inward.\cite{izidoro21}

Each of the planets may have lost a significant amount of water after accretion. High-velocity planetesimal impacts may cause water (and atmospheric) loss.\cite{schlichting15,kral18} On long timescales impacts may cause a net water gain or loss on each planet depending on the balance between stripping impacts and lower-velocity impacts.\cite{kral18} Given the radial dependence of impact speed (see Methods Sections 5 and 6), the outer planets are more likely to accumulate volatiles.\cite{kral18}  In addition, as an ultracool star Trappist-1 evolved very slowly onto the main sequence; the habitable zone swept inward such that even the planets in the present-day habitable zone spent time with much higher equilibrium temperatures.\cite{ramirez14,tian15} During this time, water in the planets' atmospheres would have been vulnerable to loss driven by ultraviolet radiation from the central star (especially for the inner planets).\cite{luger15,bolmont17}  Models of the early evolution of the Trappist-1 planets that also include the magma ocean phase\cite{kislayakova17,barth20} find that each planet may have lost a few to 20 Earth oceans of water.\cite{barth20}  Calculations that make assumptions designed to maximize water loss still find an upper limit of 25 oceans lost in the interval before planets e through h reached the habitable zone.\cite{bolmont17,barth20}  This is an order of magnitude lower than the water contents allowed by observationally-constrained internal structure modeling (Fig.~3) for each planet, compatible with a relatively water-rich formation scenario.

Our work highlights the exceptional value of multi-resonant planetary systems, only a handful of which are currently well-characterized, notably TOI-178\cite{leleu21} and Kepler-223.\cite{mills16} Not only are these systems ideally-suited for mass measurement via transit-timing variation analysis,\cite{agol21} their dynamical fragility allows us to strongly constrain their bombardment histories. Our results demonstrate that the Trappist-1 system satisfies the International Astronomical Union's (IAU) criteria for planethood.  IAU resolution B5 (see \url{https://www.iau.org/static/resolutions/Resolution_GA26-5-6.pdf}) defines a planet as an object that 1) is in orbit around the Sun, 2) is massive enough to assume hydrostatic equilibrium in a nearly round shape, and 3) has cleared the neighborhood around its orbit.  Assuming criteria 1 and 2 to be satisfied using observations of the central star\cite{gillon17} and internal structure modeling of the planets\cite{dorn18,turbet20aa}, our simulations show that the inner Trappist-1 system has been cleared of rogue material to a stringent limit.

\begin{methods}

\section*{1. Current state of knowledge of the Trappist-1 central star and the system's resonant structure}
Trappist-1 has become iconic among exoplanet systems.  The Trappist-1 planets' sizes are between 0.75 and $ 1.15 \rearth$, and their masses have been measured with a precision of 3-6\% using transit-timing variation analysis.\cite{agol21} The system's two innermost planets were discovered in 2016\cite{gillon16} and the next five in 2017.\cite{gillon17}  The central star is an ultracool dwarf star of spectral type M8 located 12 pc from the Sun.\cite{gillon16,gillon17}  A number of analyses have constrained the star's physical characteristics.\cite{burgasser17,mann19,gonzales19,ducrot20}  It has a bolometric luminosity of roughly $5.5 \times 10^{-4} L_\odot$\cite{ducrot20}, a mass of $\sim 0.09 M_\odot$ and a size of $\sim 0.12 R_\odot$.  Its age has proven hard to pin down. While one study suggest a relatively old age of $\sim 7.6$~Gyr\cite{burgasser17}, a recent study\cite{gonzales19} found analog stars that are as young as 0.5 Gyr and as old as 10 Gyr.  While this remains an interesting astronomical detective story, even a `young' age of 0.5 Gyr is far longer than our N-body simulations.  As discussed in the main text, our upper limits on late accretion are conservative, as an older system age implies a longer timescale for instability to occur (among the systems that were stable for our 10 Myr simulations).

Trappist-1's multi-resonant structure is one of its most remarkable attributes. Judging from orbital period ratios, the sequence of orbital resonances, from planets b:c outward, is: 8:5, 5:3, 3:2, 3:2, 4:3, 3:2.\cite{luger17}  Extended Data Fig. 1 shows the evolution of a number of resonant angles in a 100 million year integration of our fiducial best-fit Trappist-1 system. The innermost pair of resonances are rarely seen in simulations of resonant capture, which tend to produce first-order resonances.\cite{lee02,cresswell07,izidoro17,izidoro21,coleman19}  Papaloizou et al\cite{papaloizou18} proposed that each of these planet pairs was in fact captured into 3:2 or 4:3 resonance; i.e.\ first-order resonances. This scenario requires fast circulation of the longitudes of periastron for planets  b and c, which requires damping of their free eccentricities, which naturally happens due to tides.  This remains a plausible scenario to be confirmed with future studies combining orbital and tidal dynamics.

While 3-body (Laplace) resonance is confirmed for adjacent triplets of planets in the Trappist-1 system (e.g., see Fig.~25 in Agol et al\cite{agol21}), current constraints cannot unequivocally guarantee that 2-body resonant angles are in libration rather than circulation.  This is because 3-body resonant angles (denoted as $\phi$ in the Methods section) depend only on mean longitudes whereas 2-body angles ($\theta$ in the Methods) explicitly depend on the longitudes of periastron, which are poorly constrained by the transit times and TTVs.  We strongly suspect that 2-body angles are indeed librating, and below we show empirically that both the 2-body and 3-body resonances appear to break at roughly the same time as the planets are spread apart by dynamical interaction with rogue planetesimals (see Extended Data Fig.~3). Even if the two-body frequencies are wide of resonance, at small eccentricity the longitudes of periastron can undergo fast circulation causing 2-body angles to librate.

\section*{2. Simulations}

Our simulations started from a configuration of the Trappist-1 system derived from the most comprehensive analysis of the system to date.\cite{agol21}  We integrated the system on its own and verified that it is dynamically stable for at least 1 billion years. Table 2 lists the orbital parameters of our starting system (as well as those of three other best-fit systems that we tested; see Methods Section 5).  While the system is two-dimensional, we gave non-zero inclinations to rogue embryos and planetesimals to ensure that the systems as a whole were 3D.


\begin{table*}
\centering
\begin{tabular}{ c | c | c | c | c | c | c}
Planet & Mass  & Radius  & Semimajor & Eccentricity & Longitude of & Mean \\
 &  ($\mearth$) & ($\rearth$)& Axis $a$ (AU) & $e$ & periastron $\varpi$ ($^\circ$) & Anomaly $M$ ($^\circ$) \\
  \hline
  b & 1.3925 & 1.1174 & 0.011551 & 0.002344 & 253.61247 & 105.78489 \\
  c & 1.2943 & 1.0967 & 0.015820 &  0.001224 & 132.62793 & 54.89836 \\
  d & 0.3958 & 0.7880 & 0.02229 & 0.005045 & 202.45580 & 171.39157 \\
  e & 0.6824 & 0.9200 & 0.02930 & 0.007013 & 52.42997 & 30.97582 \\
  f & 1.0634 & 1.0448 & 0.038551 & 0.008298 & 170.04247 & 247.44087 \\
  g & 1.3464 & 1.1294 & 0.046896 & 0.003760 & 355.97714 & 87.27858 \\
  h & 0.3198 & 0.7552 & 0.061963 & 0.003571 & 172.18673 & 118.58431  \\
  \hline  
\end{tabular}
\caption*{Extended Data Table 1. Starting orbital configuration of the Trappist-1 system that were used in our fiducial simulations (referred to as 'set 1' in Methods Section 5). The planets' orbits have zero inclinations, but in our simulations rogue planets and planetesimals were given non-zero inclinations to ensure that the systems were not 2-dimensional. The system's exact configuration was drawn from the best-fit posteriors of Agol et al.\cite{agol21} }
\end{table*}

The initial conditions of our simulations were built on the assumption that one or more rogue objects was scattered inward toward the Trappist-1 planets.  Rogue objects may be thought of as analogs to the leftover planetesimals that are thought to have impacted the Earth and other terrestrial planets during the late accretion phase.\cite{day07,walker09,bottke10,raymond13,morbywood15,marchi18} If rogue objects were scattered inward from more distant orbits, they would first interact with the outer planets, and their apoastron distances would correspond to the next-most-distant planet in the system (as is the case for Jupiter-family comets in the Solar System\cite{levison94}), which is poorly constrained from observations.\cite{boss17,jontofhutter18}  In addition to the Trappist-1 planets listed in Table~2 we therefore included either one rogue planet with a randomly-chosen mass between $0.001$ and $1 \mearth$, or 1000 equal-mass rogue planetesimals whose masses totaled between $0.01$ and $3 \mearth$. The initial semi-major axis of each rogue object was randomly chosen between 0.07 and 0.1~au, with a periastron distance randomly selected to lie between 0.038 and 0.062~au (between the orbits of planets f and h), to reflect an inward-scattering origin for which the outermost planets are encountered first.\cite{levison94} We performed additional sets of simulations in which the semi-major axes of rogue planetesimals were randomly chosen between either 0.2-3~au or 0.7-1~au (see Methods Section 5). The initial inclinations of rogue objects were randomly assigned between zero and ten degrees, which matches the inclination distribution of small leftover bodies in simulations that form close-in, Trappist-1-like systems from pebble accretion and migration;\cite{izidoro21} we inherently assume this distribution to apply regardless of the exact migration scenario.\cite{terquem07,ogihara09,mcneil10,ida10,cossou14,izidoro17,ormel17,schoonenberg19,coleman19,papaloizou18,miguel20,lin21}  In simulations with 1000 planetesimals, the planetesimals felt the planets' gravity but not each others'.  Even though all planetesimals were included from the start of the simulation, the absence of mutual gravity allows us to roughly mimic the interaction of individual rogue planetesimals with the planets and estimate their cumulative effects.

Each simulation was integrated for 10 million years but stopped if an instability occurred, marked by a collision between planets.  We used the hybrid integrator in the publicly-available {\tt Mercury} integration package\cite{chambers99} with a time-step of 0.1-0.15 days, which tests showed was sufficient for long-term integration without significant numerical error.  Objects were considered to be ejected if they strayed more than 50~au from the central star, and were considered to have hit the star if they passed closer than 0.005~au, the rough size of our Sun (although note that Trappist-1 is only $\sim 0.12$ Solar radii\cite{gonzales19,ducrot20}). In terms of calculating collisions, the Trappist-1 planets were assigned their measured sizes (see Extended Data Table~1) whereas rogue planets and planetesimals were assigned sizes assuming a bulk density of 3 g~cm$^{-3}$. Collisions were treated as inelastic mergers conserving linear momentum. 

\section*{3. Characterizing and breaking orbital resonances}
We used a variety of resonant angles to characterize the planets' resonances.  The Trappist-1 resonant chain involves a series of mean motion and Laplace resonances.\cite{luger17}  From inside to out, the planet pairs are in or near 8:5, 5:3, 3:2, 3:2, 4:3, and 3:2 mean motion resonance.  In addition, each set of three planets is in Laplace resonance.

For each planet pair in mean motion resonance $p+q : p$, resonant angles $\theta_i$ are of the form
\begin{equation}
\theta_{1,2} = (p+q) \lambda_1 - p \lambda_2 -q \varpi_{1,2} \\
\label{eqn:mmr}
\end{equation}
\noindent where $\varpi$ are longitudes of pericenter, $\lambda$ are mean longitudes (equal to the sum of $\varpi$ and the mean anomalies), and subscripts 1 and 2 refer to the inner and outer planet, respectively.  Resonant angles measure the angle between the planets at conjunction; if any angle librates rather than circulates, then the planets are in resonance.  Different resonances have different numbers of resonant arguments, involving various permutations of the final terms in Eq.~1.  The order of a resonance is given by $q$. 

In our analysis we calculated the relevant $\theta$ values for each planet pair. We focused on the outermost pair of planets, as we found that their 3:2 resonance was often the first to break. We also calculated 3-body (Laplace) resonant angles. There are five such angles in the system, one for each adjacent triplet of neighboring planets. Using subscripts appropriate for the Trappist-1 planets, these are (see Table 1 from Luger et al\cite{luger17}):
\begin{eqnarray}
\phi_1 = 2 \lambda_b - 5 \lambda_c + 3 \lambda_d
\nonumber \\
\phi_2 = \lambda_c - 3 \lambda_d + 2 \lambda_e
\nonumber \\
\phi_3 = 2 \lambda_d - 5 \lambda_e + 3 \lambda_f
\nonumber \\
\phi_4 = \lambda_e - 3 \lambda_f + 2 \lambda_g
\nonumber \\
\phi_5 = \lambda_f - 2 \lambda_g + \lambda_h
\end{eqnarray}


Extended Data Fig.~1 demonstrates the multi-resonant nature of the initial conditions for the Trappist-1 system that we used in our simulations. It shows the evolution of different mean-motion and Laplace resonant angles over a 100 million year integration of the fiducial system (orbital parameters in Table~2 of the main text).  Each of these angles is in a state of libration, meaning that its evolution is constrained to a limited range of possible values. This indicates a repeated alignment of the planets during their conjunctions, and is the definition of orbital resonance.\cite{murraydermott99}. 

\begin{figure*}
  \begin{center} 
\leavevmode \epsfxsize=12cm\epsfbox{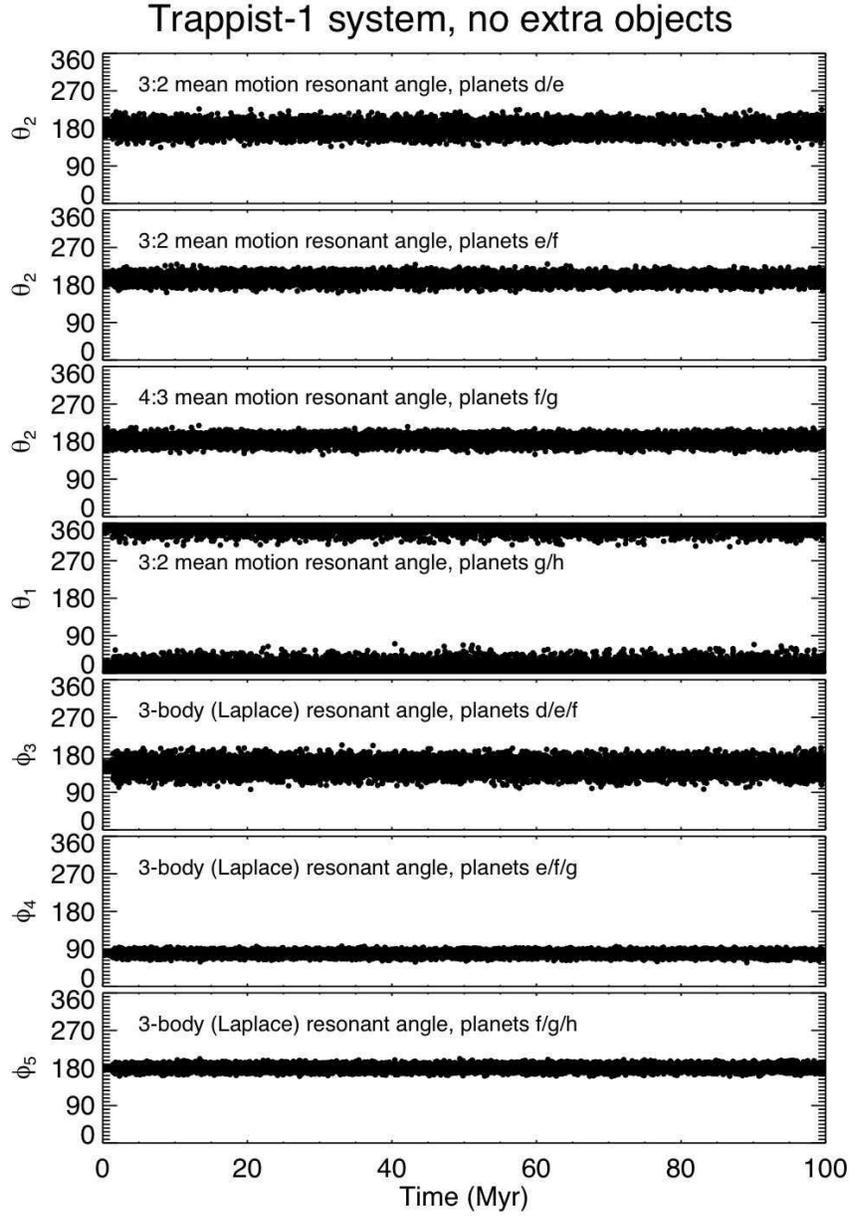}
    \caption*{{\bf Extended Data Fig. 1. }Evolution of resonant angles for our fiducial Trappist-1 system with no added objects.  The top four panels each show one angle that characterizes the mean motion resonance between a pair of neighboring planets, as defined in Equations 1 and 2.  The bottom three panels show angles that characterize the 3-body Laplace resonances between triplets of planets, as defined in Eq.~3.
        } 
     \label{fig:sim1}
    \end{center}
\end{figure*}

Extended Data Fig.~2 presents three examples of the evolution of resonant behavior from simulations with 1000 rogue planetesimals.  In the simulation interacting with the lowest mass in planetesimals (a total of $0.0213 \mearth$; left panel), the planets maintained their resonant configuration with no noticeable difference compared with the fiducial system from Extended Data Fig.~1.  When the mass in planetesimals was doubled to $0.0429 \mearth$ (center panel), the resonances were strained.  The mean motion resonances between planets d and e, e and f, and f and g were maintained (top three center panels). However, the 3:2 resonance between planets g and h was compromised, as the resonant angle $\theta_2$ no longer shows clear libration but rather circulates.  However, all three 3-body resonances were maintained, as shown by the continued libration of the Laplace angles $\phi_3$, $\phi_4$, and $\phi_5$ (bottom three center panels).  While the exact center of libration underwent shifts during the integration, we consider the resonance to have been maintained to a sufficient degree that, if the system were detected and observed over a relatively short temporal arc, it would appear to be in resonance.  Resonances were completely disrupted in the system interacting with just a slightly higher mass in planetesimals, $0.0645 \mearth$ (right panel).  In that case, all of the resonant angles associated with mean-motion and Laplace resonances transition from libration to circulation within roughly 1 Myr of the start of the integration.  While the system remained stable for the 10 Myr length of the integration, it is possible that it would be unstable on longer timescales.\cite{tamayo17}

\begin{figure*}
  \begin{center} 
  \leavevmode \epsfxsize=5.15cm\epsfbox{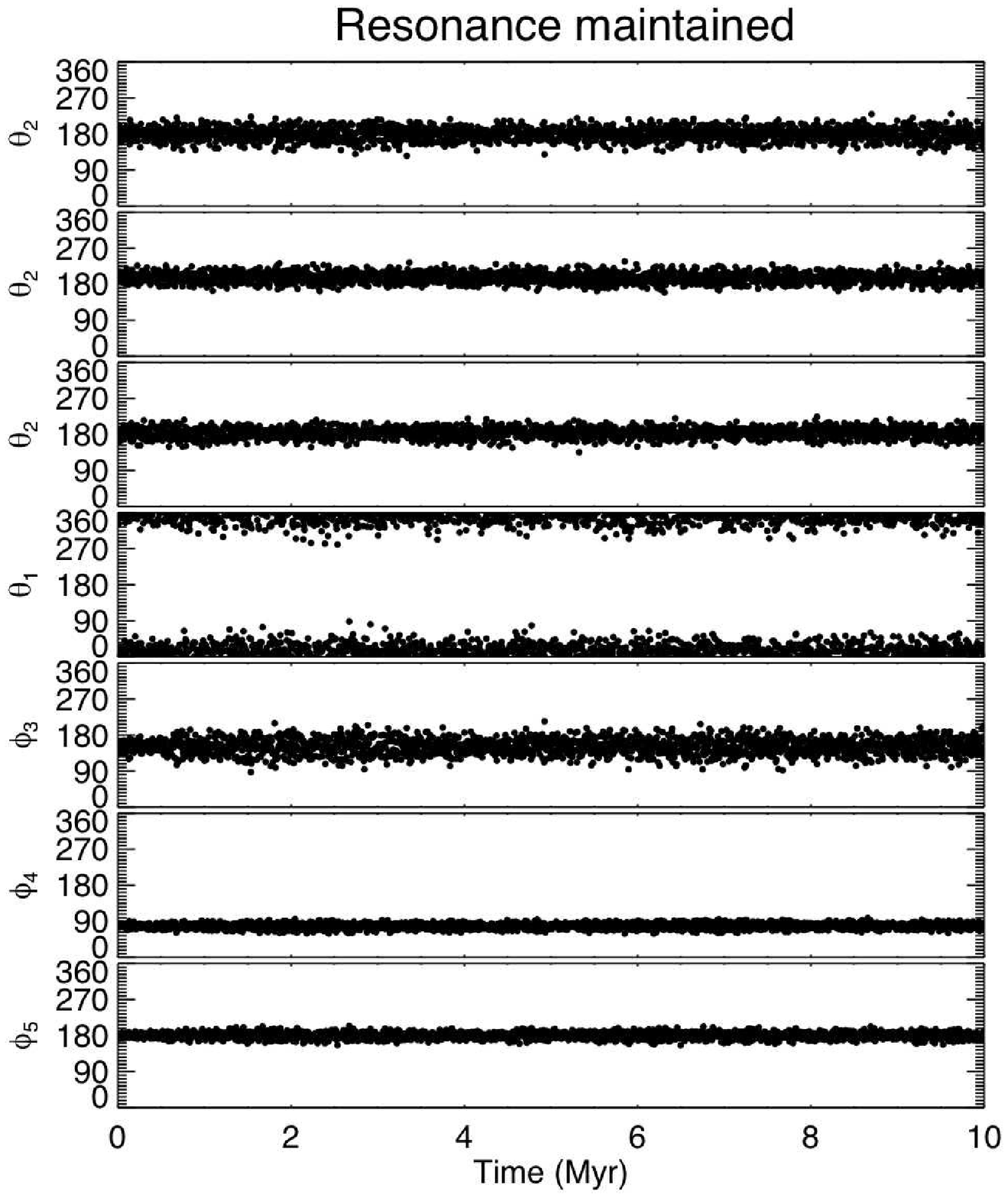}
  \leavevmode \epsfxsize=5.15cm\epsfbox{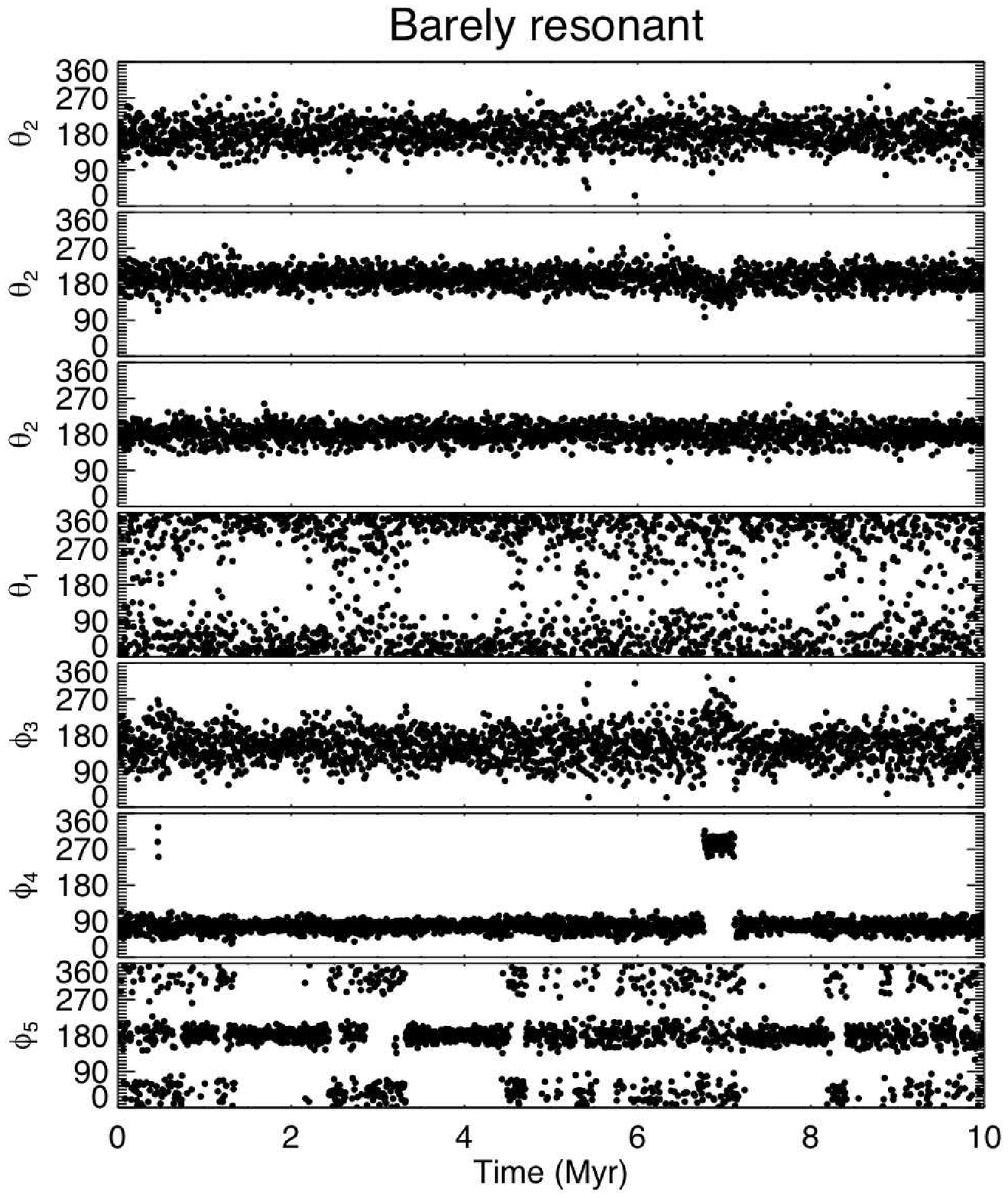}
  \leavevmode \epsfxsize=5.15cm\epsfbox{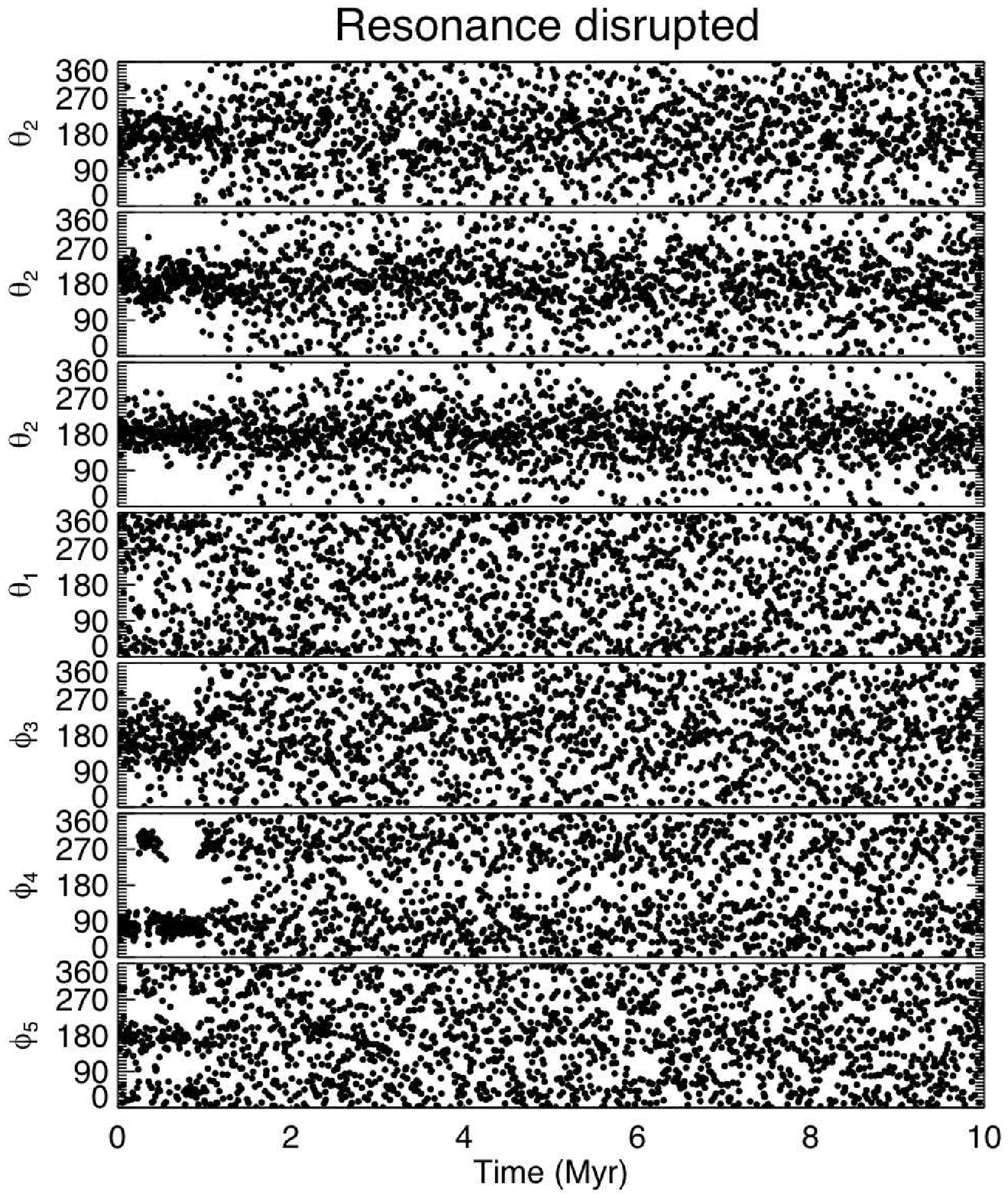}
    \caption*{{\bf Extended Data Fig. 2. } Evolution of the resonant angles for three simulations that included our fiducial Trappist-1 system and 1000 rogue planetesimals.  Each system was stable for the full 10 Myr simulation.  The resonant angles that are shown are the same ones presented in Extended Data Fig.~1.  The total mass in planetesimals for each system was $0.0213 \mearth$ (left panel; resonance maintained), $0.0429 \mearth$ (middle panel; resonance barely maintained), and $0.0645 \mearth$ (right panel; resonance disrupted). 
        } 
     \label{fig:examples}
    \end{center}
\end{figure*}

Extended Data Fig.~3 shows the disruption of the resonances in a simulation in which our fiducial Trappist-1 planets interact with a rogue planetesimal swarm of $0.0967 \mearth$.  The widening the outer planets' orbits is small, and the changes in the planets' semimajor axes can be measured in ten-thousandths of an astronomical unit.  The resonances were maintained for the first several thousand years.  The 3:2 mean motion resonance between planets g and h was disrupted once the relative separation of the two planets increased by roughly 0.0002~au.  The 4:3 resonance between planets f and g was likewise destroyed when those planets' relative separation increased by $\sim 0.00015$~AU, which happened at a later time than the disruption of the 3:2 resonance between planets g and h because of the different rates of relative drift.  The $\phi_5$ Laplace angle for planets f/g/h was disrupted at roughly the same time as the 3:2 mean motion resonance between planets g and h, and the $\phi_4$ Laplace angle for planets e/f/g was disrupted on a similar timeframe.

 \begin{figure*}
  \begin{center} 
  \leavevmode \epsfxsize=12cm\epsfbox{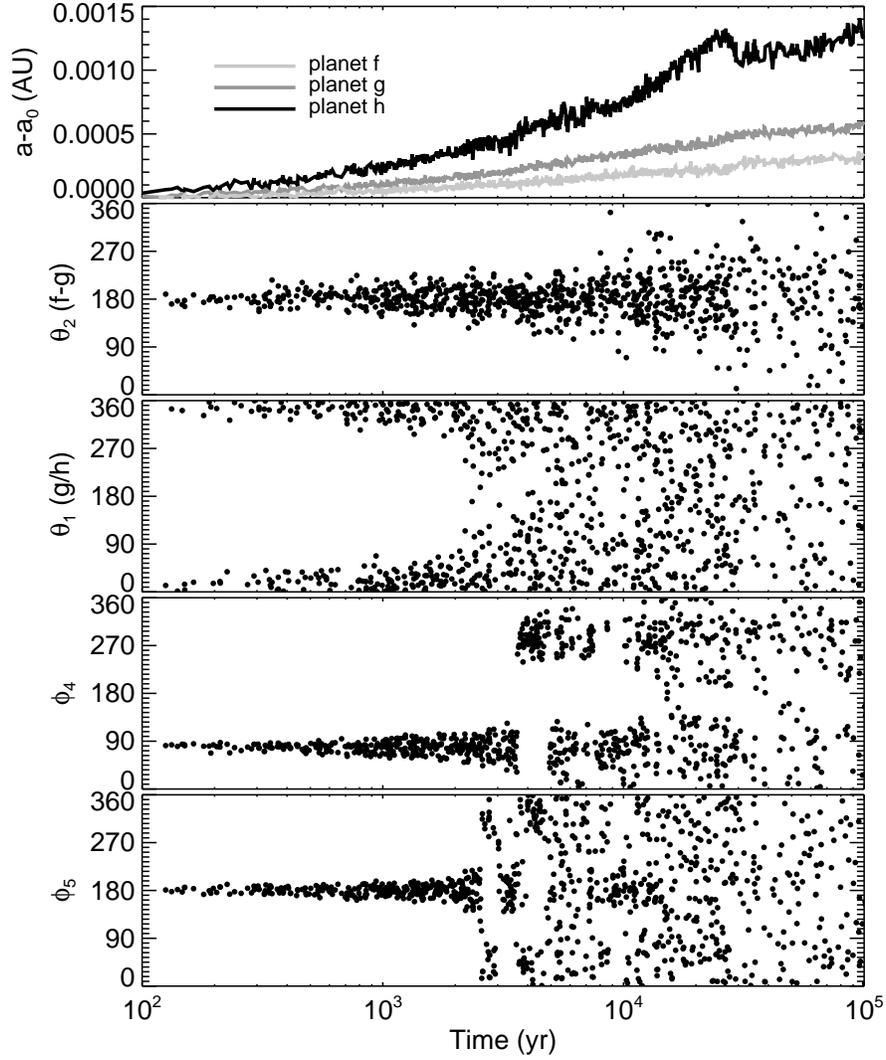}
    \caption*{{\bf Extended Data Fig. 3. } Disruption of the resonances in the outer parts of the Trappist-1 system.  In this simulation, the fiducial system of planets interacted with 1000 planetesimals containing a total of $0.0967 \mearth$.  The top panel shows the increase in the semimajor axis of the orbits of planets f, g and h.  The other panels show the angles associated with mean-motion and Laplace resonances between the planets. 
        } 
     \label{fig:disrupt}
    \end{center}
\end{figure*}

To understand why a given degree of planetesimal scattering disrupts orbital resonances we must first calculate the width of a resonance, and then determine how much mass a planet must scatter to shift its orbital radius enough to leave the resonance.  Following the method of Hadden \& Lithwick\cite{hadden18} (see their equations 5, 11 and 15) we estimate the width of the 3:2 mean motion resonance between planets g and h to be $\Delta a \approx 2.4 \times 10^{-4}$~AU.  This is in good agreement with the evolution of the system shown in Extended Data Fig.~3, in which the 3:2 resonance between planets g and h was indeed disrupted once their relative semimajor axes increased by $\sim 2 \times 10^{-4}$~au.  

How much mass in planetesimals must be scattered to remove a pair of planets from resonance?  The interactions between the planets and rogue planetesimals involve the exchange of orbital angular momentum, $L$.  The angular momentum of a planet of mass $M_p$ is defined as:
\begin{equation}
L = M_p \sqrt{G M_\star a} \left(1 - cos(i) \sqrt{1 - e^2}\right),
\end{equation}
\noindent where $G$ is the gravitational constant, and $i$ is the orbital inclination.  

If a planetesimal of mass $m$ transfers all of its angular momentum to a planet of mass $M_p$, that planet's orbit will shift by $\delta a/a \approx 2 m/M_p$ (neglecting terms related to eccentricity and inclination, and assuming a common semimajor axis for the planetesimal and planet).  Based on this calculation it should require $\sim 0.005-0.01 \mearth$ in order to move planet h by the width of its common 3:2 resonance with planet g. Indeed, this is roughly the minimum mass in a single rogue planet that destroys the system's resonant structure (see Fig.~1 in the main text and Extended Data Fig.~5).  However, a total mass in rogue planetesimals of $\sim 0.05 \mearth$ is needed to break the resonance. The reason for this is that planet h does not orbit its star in isolation. Planet h does not take all of a typical planetesimal's angular momentum but rather scatters the planetesimal inward to interact with planet g.  The back-reaction of the scattering causes planet h's semimajor axis to shift slightly outward. Subsequently, the same process is repeated as the planetesimal interacts with planet g, causing its semimajor axis to also shift slightly.  Thus, the planets migrate outward together, as is clearly seen in the example from Extended Data Fig.~3.  This process is not perfectly balanced, and in time the planets' orbits do drift apart but this mutual drift is significantly slower than each of their individual drift rates.  Thus, the planets actually drift several times their resonant width outward before they separate enough to leave the resonance. For this reason it takes substantially more than $0.01 \mearth$ to disrupt the resonance.  Hence, the system's resonant structure requires less mass to be disrupted by a single rogue planet than by a swarm of rogue planetesimals.  

Our upper limits on the mass with which the Trappist-1 system could have interacted are conservative. Once a system loses its multi-resonant structure or goes unstable, it is irreversible: the system cannot regain its resonances (after the end of the gaseous disk phase -- tidal evolution is not strong enough; see Methods Section 7) and it is forever inconsistent with the present-day system. Yet a fraction of systems that remained in resonant chains for the duration of our 10 Myr simulations could lose their resonances or become unstable on longer timescales. A hypothetical system that loses its resonant structure on a longer timescale could not interacted with more total mass than our upper limits.  By using non self-interacting particles, our simulations constrain the cumulative effect of a given mass in rogue planetesimals. For example, a system that interacted with $0.02 \mearth$ in rogue planetesimals during the first 10 Myr of its history would retain its multi-resonant structure.  Given the upper limit of $0.05 \mearth$ from Fig.~1 in the main text, this system could not interact with more than an additional $0.03 \mearth$ on longer timescales without losing its resonances.

\section*{4. Testing the effect of our initial orbital distribution of rogue objects}

Our choices for the orbital distribution of rogue planets and planetesimals were motivated by the well-studied dynamics of Jupiter-family comets\cite{levison94} and of leftovers from terrestrial planet formation\cite{raymond13}, and from the orbital distribution of analogous leftover material in simulations of the formation of close-in systems like Trappist-1.\cite{izidoro21}  Yet it is worth questioning these choices, and exploring how robust our conclusions are to other plausible populations of rogue objects (for a discussion see Kral et al\cite{kral18}).  

We therefore performed two additional sets of simulations.  In each additional set the orbits of the planets were unchanged. Rogue planetesimals were included with the same inclination distribution (randomly selected from zero to 10 degrees -- derived from simulations of the formation of Trappist-1-like systems\cite{izidoro21}) and periastron distributions (randomly chosen between 0.038 and 0.062~au) as in our fiducial set.  The reason for keeping these parameters identical was to isolate the importance of the rogue objects' semimajor axes.

The new sets of simulations were set up as follows:
\begin{itemize}
    \item A set of 50 simulations in which the semimajor axes of rogue objects were randomly chosen between 0.2 and 0.3~au, roughly three times larger than in our fiducial set.  50 simulations were performed with a total mass in rogue planetesimals evenly chosen in the range $0.01-1\mearth$.
    \item In the second set, the semimajor axes of rogue objects were randomly chosen between 0.7 and 1~au, a full order of magnitude larger than in our fiducial set. 50 simulations were performed with a total mass in rogue planetesimals evenly chosen in the range $0.01-1\mearth$.
\end{itemize}

All other integration parameters (e.g., timestep, ejection radius, stellar mass; see Methods Section 2) were the same as for our fiducial set of simulations. 

\begin{figure*}
  \begin{center} 
    \centering 
    \epsfxsize=8cm\epsfbox{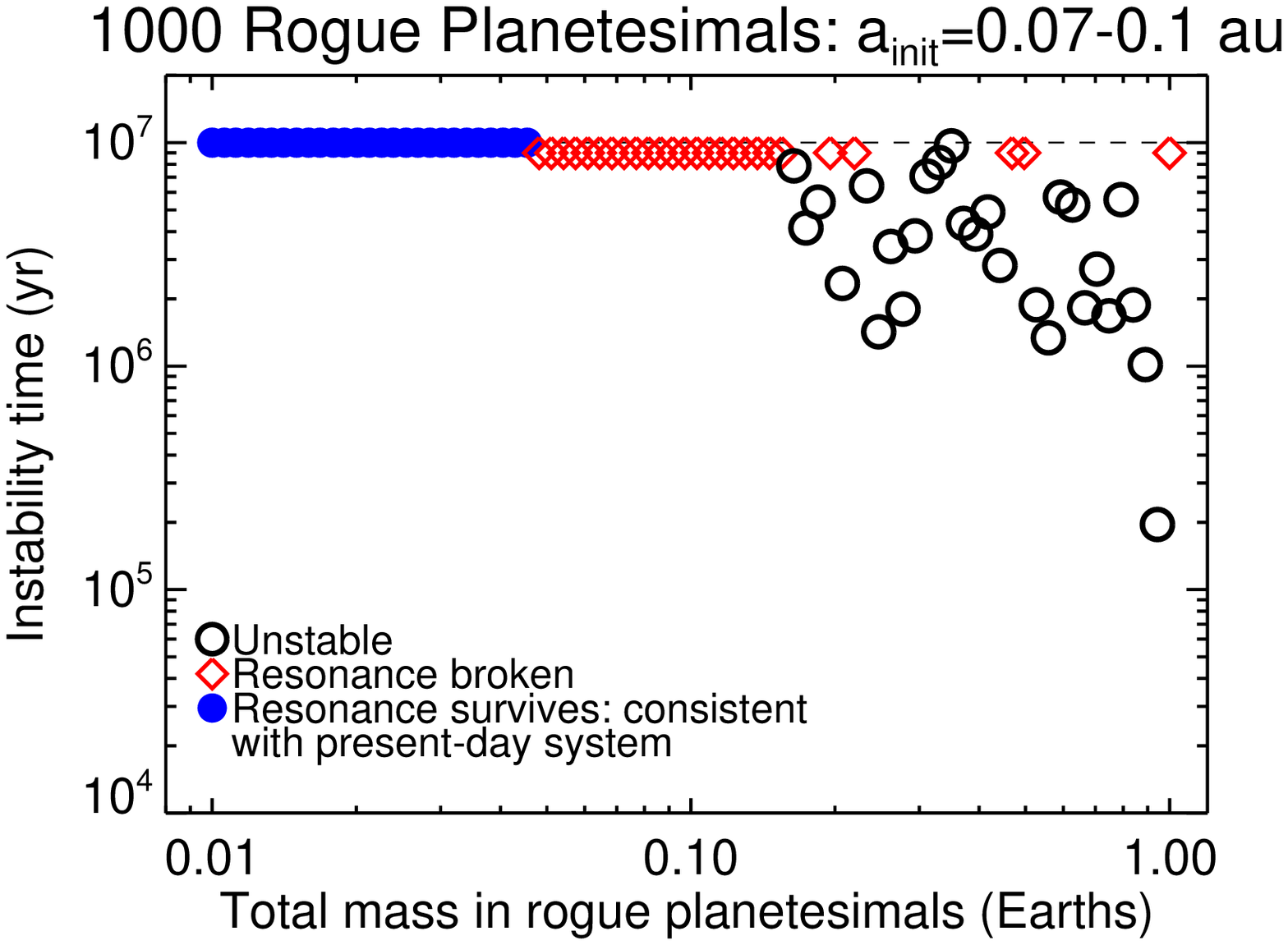}      
\epsfxsize=8cm\epsfbox{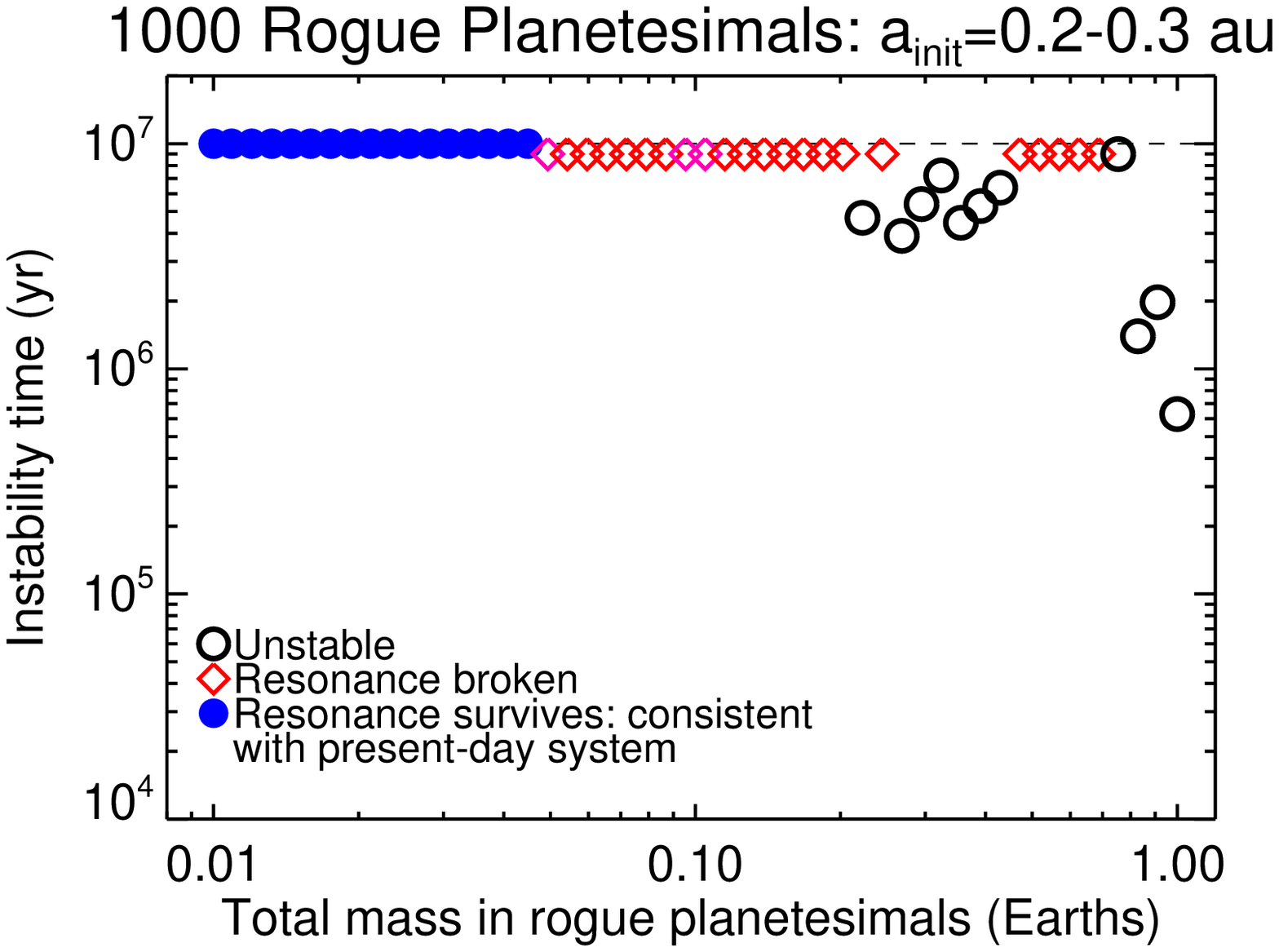}     \\
\epsfxsize=8cm\epsfbox{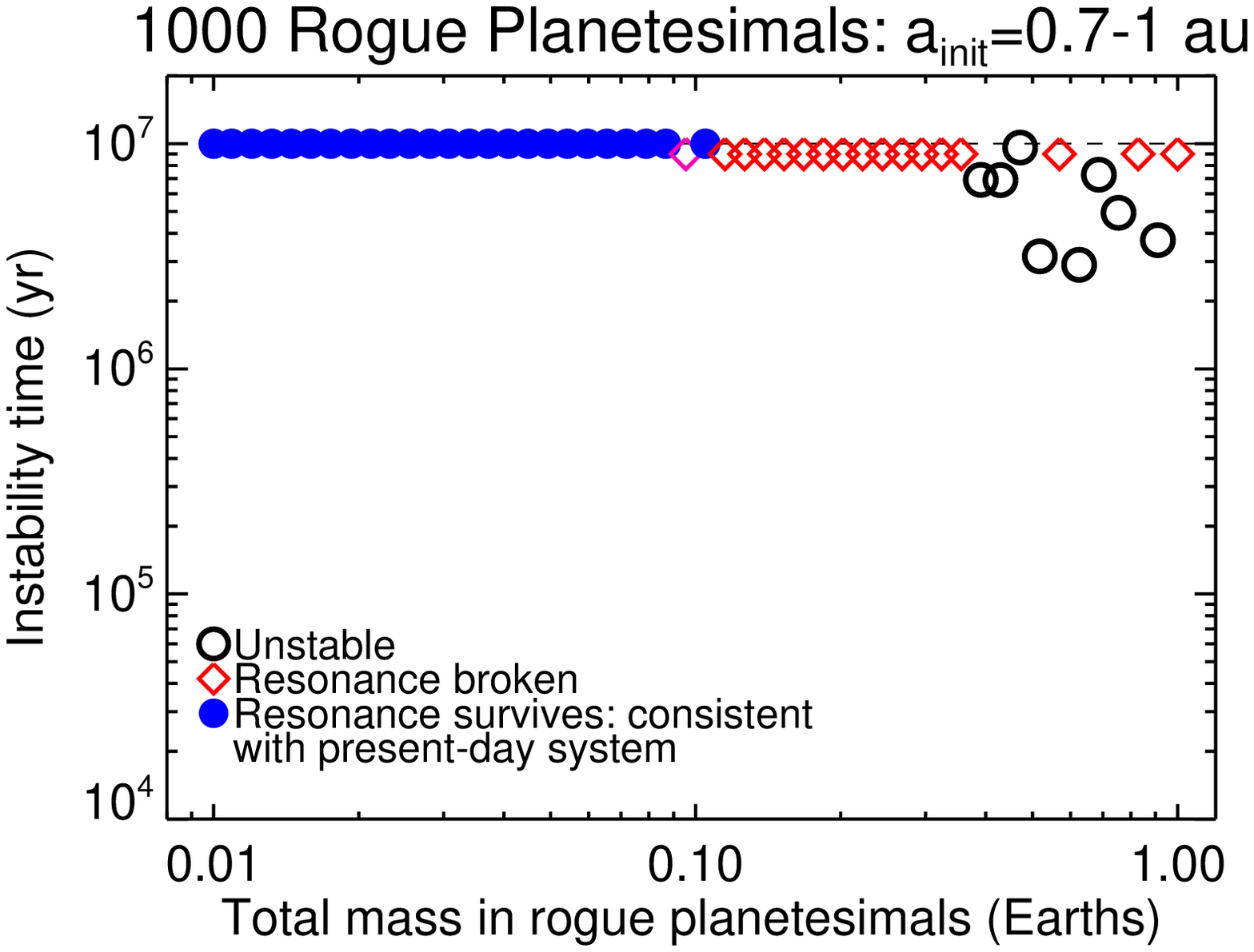} 
    \epsfxsize=8cm\epsfbox{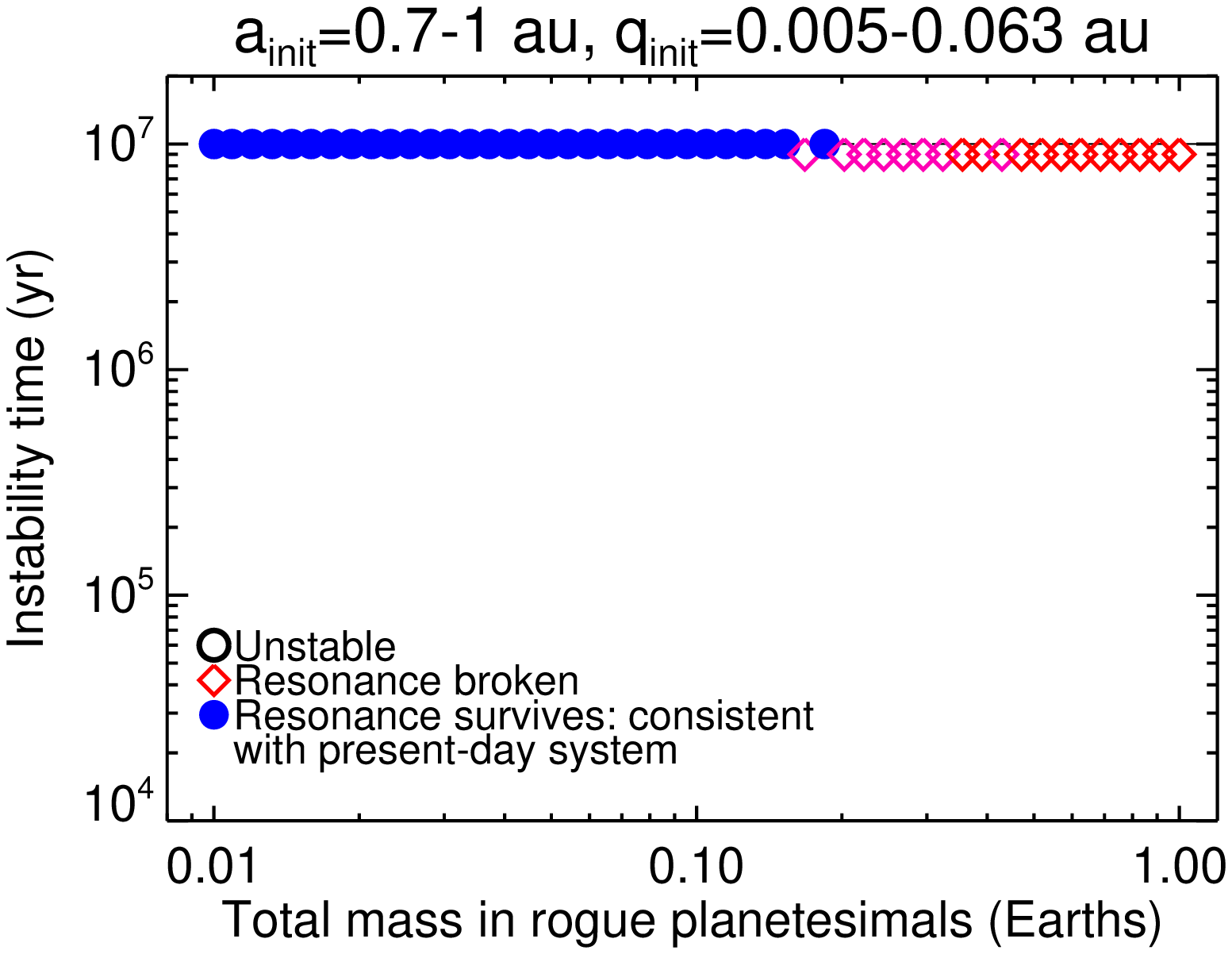}     

    \caption*{{\bf Extended Data Fig. 4. }Effect of the semimajor axis distribution of rogue planetesimals. The top left panel shows our the subset of our fiducial batch of simulations with 1000 rogue planetesimals with a total mass in rogue planetesimals between 0.01 and 1 Earth mass.  The top right and bottom left panels show the sets of simulations with $a_{rogue} = 0.2-0.3$~au and $a_{rogue} = 0.7-1$~au, and the same periastron distribution as in the fiducial set (with orbits initially crossing those of only the outer planets). The bottom right panel shows a set of simulations in which the rogue planetesimals were on extremely eccentric orbits, with semimajor axes $a_{rogue} = 0.7-1$~au and periastron distance $q_{rogue}$ randomly chosen to sample a logarithmic distribution in the range $0.005-0.06$~au. As in Fig.~1 in the main text, blue filled circles indicate systems that remained in resonant chains, red diamonds those that remained stable for 10 Myr but lost their resonant structure, and black empty circles those that underwent dynamical instabilities. }
     \label{fig:peritest}
    \end{center}
\end{figure*}

\begin{table*} 
\centering
\begin{tabular}{ c | c | c | c | c | c}
Planet & Orbital radius & Max. bombard. & Max. bombard. & Max. bombard.  & Max. bombard. \\
 & (AU) & mass ($\mearth$) & mass ($\mearth$) & mass ($\mearth$) & mass ($\mearth$)  \\
 & & [0.07-0.1~au] & [0.2-0.3~au] & [0.7-1~au]  & [0.7-1~au] \\
& & & & & [super-ecc.] \\
  \hline			
  b & 0.0115 & 0.00035 & 0.00013 & 0.0001 & 0.0082 \\
  c & 0.0158 & 0.0015 & 0.0005 & 0.00023  & 0.0073 \\
  d & 0.0223 & 0.0015 & 0.00047 & 0.0003  & 0.0033 \\
  e & 0.0293 & 0.0034 & 0.00098 & 0.00087 & 0.0044 \\
  f & 0.0385 & 0.0078 & 0.0038 & 0.003    & 0.0063 \\
  g & 0.0469 & 0.017 & 0.01  & 0.0083     & 0.0074 \\
  h & 0.0620 & 0.014 & 0.013  &  0.011    & 0.0045 \\
  \hline  
\end{tabular}
\caption*{Extended Data Table 2. Upper limits on the total mass in planetesimals that could have collided with each of the Trappist-1 planets (assuming perfect mergers), for four sets of initial orbital distributions for the rogue planetesimals.  The first three columns share the same periastron distribution, with starting semimajor axes of 0.07-0.1~au (our fiducial set), 0.2-0.3~au, and 0.7-1~au.  The fourth column is from the set in which rogue planetesimals have starting semimajor axes of 0.7-1~au and super-eccentric orbits, with starting periastron distances of $0.005-0.063$~au. For each set we used the maximum total initial planetesimal mass that preserved the resonant structure of the system: $0.05 \mearth$ for $a_{rogue} = 0.07-0.1$~au and $a_{rogue} = 0.2-0.3$~au, $0.1 \mearth$ for $a_{rogue} = 0.7-1$~au with the standard periastron distribution, and $0.018 \mearth$ for $a_{rogue} = 0.7-1$~au with super-eccentric orbits (see Extended Data Fig.~4.)}
\end{table*}

Extended Data Fig.~4 compares the outcomes of these new sets of simulations with our fiducial set. There is no clear difference in the figure between simulations with $a_{rogue} = 0.07-0.1$~au and those with $a_{rogue} = 0.2-0.3$~au (for the same periastron distribution).  In both cases a total mass in planetesimals larger than $\sim 0.05 \mearth$ universally disrupts the system's multi-resonant architecture. In contrast, simulations with $a_{rogue} = 0.7-1$~au the critical mass in planetesimals is increased to $\sim 0.1 \mearth$.

The central difference between the different sets of simulations is the fraction of planetesimals that is ejected from the system rather than accreted by a planet.  Only considering simulations that retained their multi-resonant structure, 7.1\% of planetesimals in our fiducial simulations (with $a_{rogue} = 0.07-0.1$~au) were scattered beyond 50~au and considered to be ejected from the system.  This fraction increased dramatically in the new sets of simulations: 41.6\% of planetesimals in simulations with $a_{rogue} = 0.2-0.3$~au were ejected, and 76.5\% of planetesimals were ejected in simulations with $a_{rogue} = 0.7-1$~au.  

The tendency for increased planetesimal ejected for larger $a_{rogue}$ can be understood from simple dynamical considerations (see also the discussion in Section 3.2 of Kral et al\cite{kral18}). To be ejected, rogue objects must be gravitationally scattered by the planets, and the maximal change in velocity that a planet can impart is the escape speed from its surface.\cite{wyatt17} Orbital energy scales with the apoastron distance as $\sim G M_\star/Q$, such that a smaller velocity kick is required to eject an object on a more distant orbit.  For an Earth-mass, Earth-sized planet, comets can be efficiently unbound for $Q$ larger than roughly 0.2~au (see Eq. 3 in Kral et al\cite{kral18}).  Our simulations match this expectation, as rogue objects are preferentially accreted in our fiducial simulations with $a_{rogue} = 0.07-0.1$~au, preferentially ejected in simulations with $a_{rogue} = 0.7-1$~au, and are roughly evenly-split between accretion and ejection in simulations with $a_{rogue} = 0.2-0.3$~au.

Compared with the fiducial set, simulations with larger values of $a_{rogue}$ allow for {\em less} late accretion on each planet (Extended Data Table~2).  These upper limits were calculated by tabulating the impact rates on each of the planets in simulations that preserved their multi-resonant structure, then multiplying by the maximum allowed total planetesimal mass.  The difference between the fiducial set and the new sets is largest for the inner planets.  This is because the initial periastron distances of planetesimals only crossed those of planets f through h; close encounters with the planets are required for a planetesimal's orbit to shrink its close approach distance.  Yet for planetesimals with large $a_{rogue}$ such encounters often lead to ejection such that, compared with planetesimals with small $a_{rogue}$, a smaller fraction of objects enters the innermost parts of the system to potentially collide with the inner planets.

These sets of simulations demonstrate the robustness of the upper limits on late accretion in the main paper that were derived using our fiducial initial conditions. While other orbital distributions of rogue planetesimals may imply somewhat higher values for the total mass in planetesimals that could have interacted with the planets (Extended Data Fig.~4), the total accreted mass is universally lower (first three columns in Extended Data Table~2).  In addition, the higher impact speeds of more distant planetesimals may often lead to erosion rather than accretion.\cite{kral18}

We performed another set of simulations in which the planetesimals were on even more extreme orbits.  The planetesimals' semimajor axes were randomly chosen in the range $a_{rogue} = 0.7-1$~au, and their periastron distances were chosen following a logarithmic distribution between 0.005~au and 0.063~au; the planetesimals' eccentricities thus spanned from 0.91 to 0.995.  As before, we performed 50 simulations varying the total mass in rogue planetesimals between $0.01$ and $1 \mearth$.  

Extended Data Fig.~4 (bottom right panel) presents the results of this set of simulations with rogue planetesimals on extreme orbits, similar to those modeled by Kral et al.\cite{kral18}  The most striking features are the absence of unstable orbits and the larger critical mass in rogue planetesimals below which the systems maintained their multi-resonant structure. This can be understood as a consequence of the fact that planetesimals on extremely high eccentricities carry very little orbital angular momentum such that a relatively large mass is needed to move the planets' orbits out of resonance.

Despite the increased mass with which the planets interacted, the amount of late accretion did not change dramatically (Extended Data Table 2).   Part of this was a consequence of ejection: as in other sets of simulations with large semimajor axes, a large fraction (77\%) of planetesimals was ejected, reducing the reservoir of potential impactors.  For planets f, g, and h the upper limits on late accretion in this set of simulations is significantly lower than in our fiducial set of simulations, with maximum late accreted masses comparable to Earth's value of $\sim 0.005 \mearth$.\cite{day07,walker09}  For planet e the upper limit on late accretion was similar in both cases ($\approx 0.003-0.004 \mearth$).  For planets b, c, and d, however, the upper limits were significantly higher in simulations with super-eccentric rogue planetesimals, with values up to $0.008 \mearth$ (for planet b; Extended Data Table 2).  This is a consequence of the much higher number of planetesimals initially placed on orbits crossing the orbits of the inner planets (due to the logarithmic distribution of periastron distances), and the fact that impact rates are higher for the inner planets.\cite{kral18}

One might wonder whether a population of planetesimals at even higher semimajor axis could in principle increase our derived upper limits on late accretion (and perhaps on a longer timescale if the semimajor axes are sufficiently large).  We do not think this is the case. In our extra sets of simulations the upper limits on accretion did not increase for any planet (see Table 1).  (The exception is the set of simulations with super-eccentric orbits from Table 2 in which the increase for planets b-d comes from a change in the initial periastron distribution of rogue planetesimals and not from an increase in the semimajor axes). Further increasing the semimajor axes of rogue planetesimals would further decrease their specific orbital angular momentum and increase the upper limits on mass with which the planets could have interacted, but this effect would be counter-balanced by the much increased ejection efficiency and the lower accretion efficiency, as quantified by Kral et al\cite{kral18}.  Of course, the origin of a massive flux of objects crossing the planets’ orbits is a separate issue that was also addressed by Kral et al.\cite{kral18}  

Upper limits on late water delivery from rogue planetesimals may be affected if planetesimals originating farther from the star have higher water contents. Comets are generally assumed to be half water by mass,\cite{ahearn11} although the bulk water contents of the best-studied comet (67P/Churyumov-Gerasimenko) is roughly constrained to $\sim 25\%$ or lower (see Fig 5b in Choukroun et al\cite{choukroun20}).  If planetesimals from 0.7-1~au have bulk water contents of 25-50\%, then the corresponding upper limits could modestly increase relative to our fiducial case -- for instance, for planets g and h in the simulations presented in Table~1 and for all of the planets in the simulation with extreme eccentricities (Table 2).  However, the higher impact velocities of planetesimals originating at 0.7-1~au could reduce the actual delivery rate of water through imperfect accretion and impact volatilization.\cite{kral18,schlichting15}

\section*{5. Testing the effect of our chosen configuration for the Trappist-1 planets' orbits}

One might worry that our results were affected by the exact choice of initial conditions, in other words by the particular ``best-fit'' system that we chose. Of course, this is just one of a population of configurations that are consistent with the current observational data set.  To address this concern, we ran similar batches of simulations with rogue planets and planetesimals in three other best-fit configurations that match all current constraints for the Trappist-1 system.\cite{agol21} Long-term simulations show that each of these three configurations is stable in isolation for at least 1 billion years. 

For each of these three systems (full configurations listed in Table 1 of the Supplementary Information), we ran both two sets of simulations that were very similar to our fiducial set:
\begin{itemize}
    \item A set of 500 simulations that included a single rogue planet with mass in the same range as for our fiducial set  ($M_{rogue} = 0.001-1 \mearth$) and the same orbital distribution as described above.  
    \item A set of 50 simulations that included 1000 planetesimals, with a total mass of $0.01-1\mearth$ and the same orbital distribution as for the fiducial simulations.
\end{itemize}

The integration method was the same as in the main set of simulations (see Methods), utilizing the hybrid {\tt Mercury} integrator.\cite{chambers99} 


\begin{figure*}
  \begin{center} 
    \leavevmode \epsfxsize=6.0cm\epsfbox{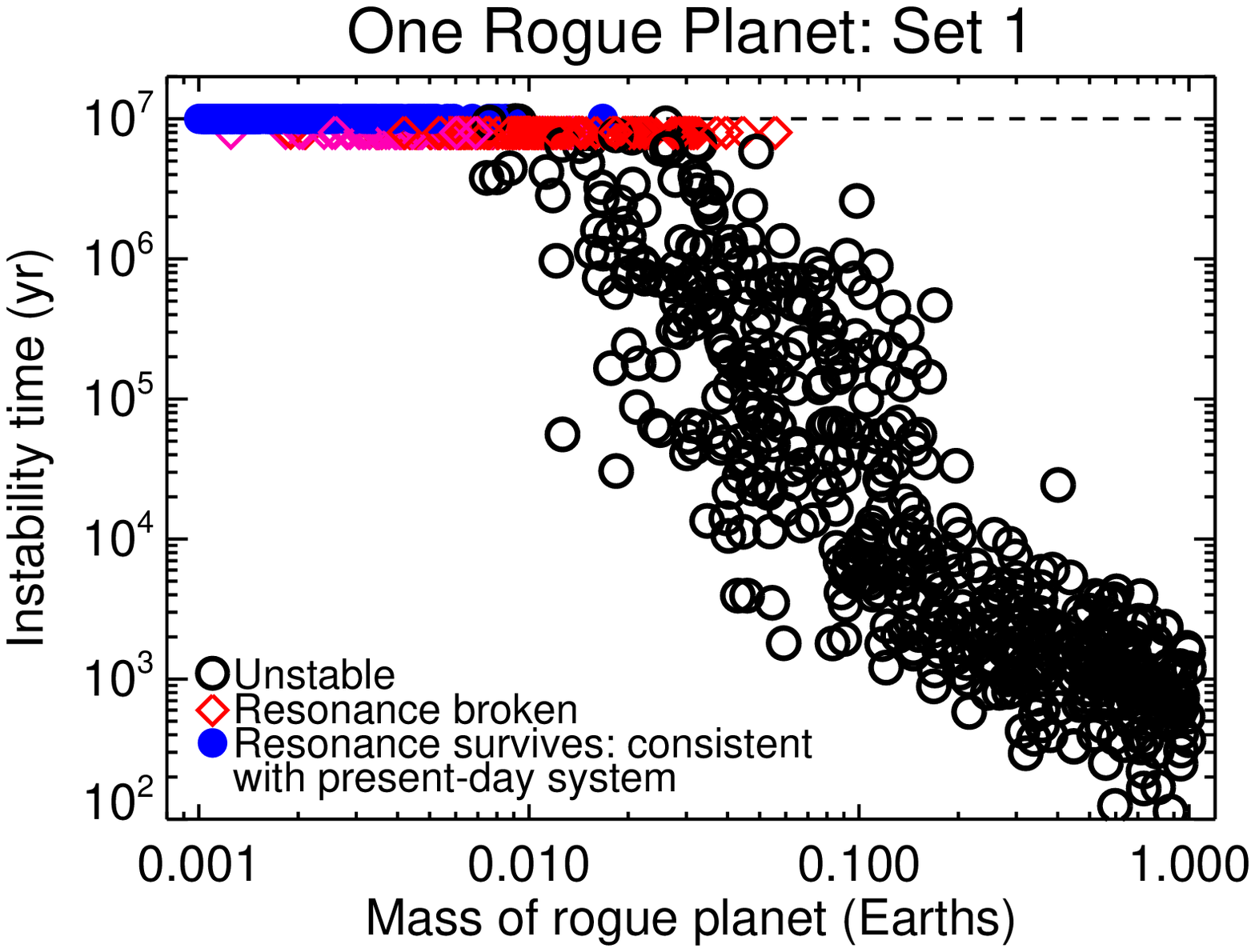} 
  \leavevmode \epsfxsize=6.0cm\epsfbox{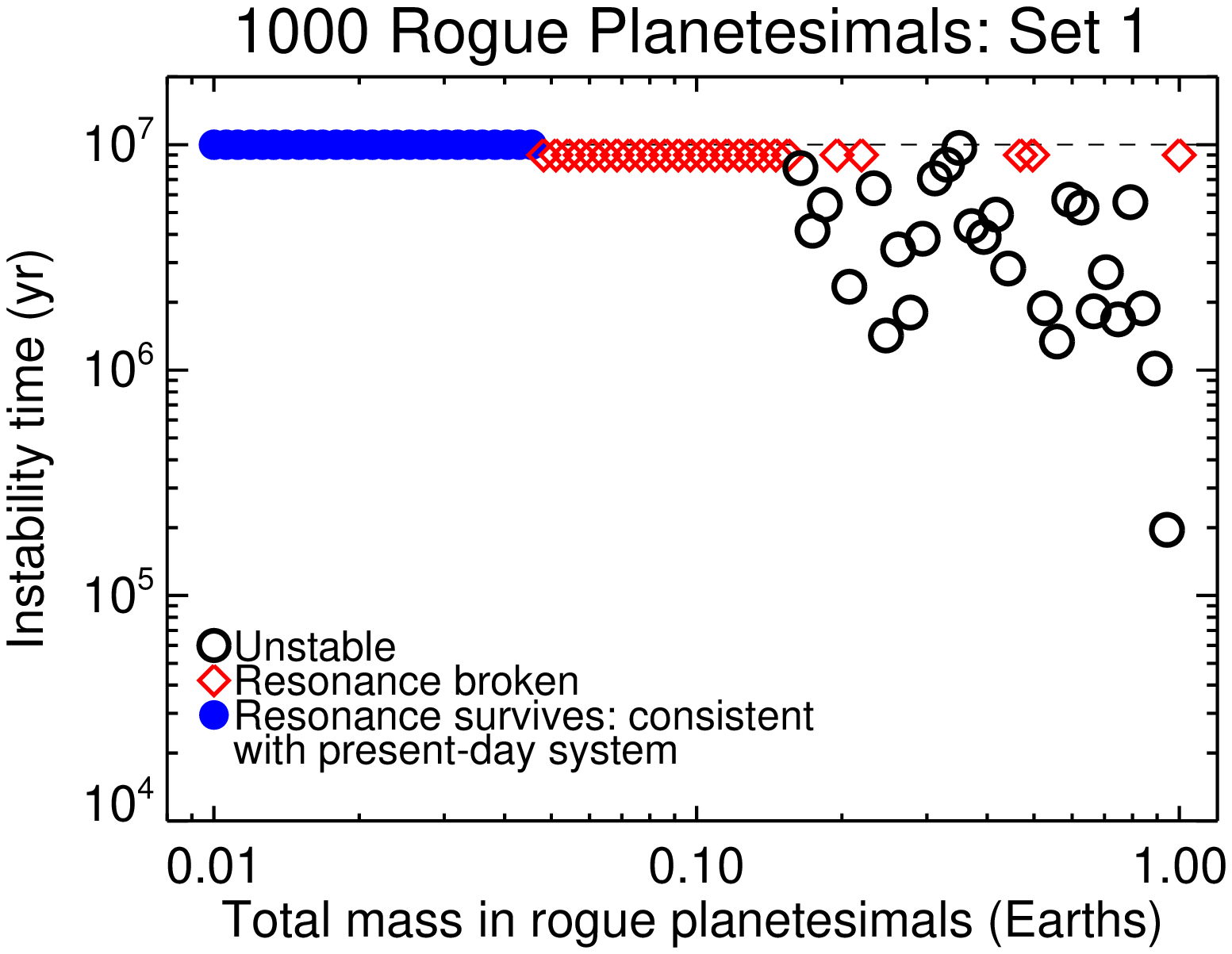} \\
      \leavevmode \epsfxsize=6.0cm\epsfbox{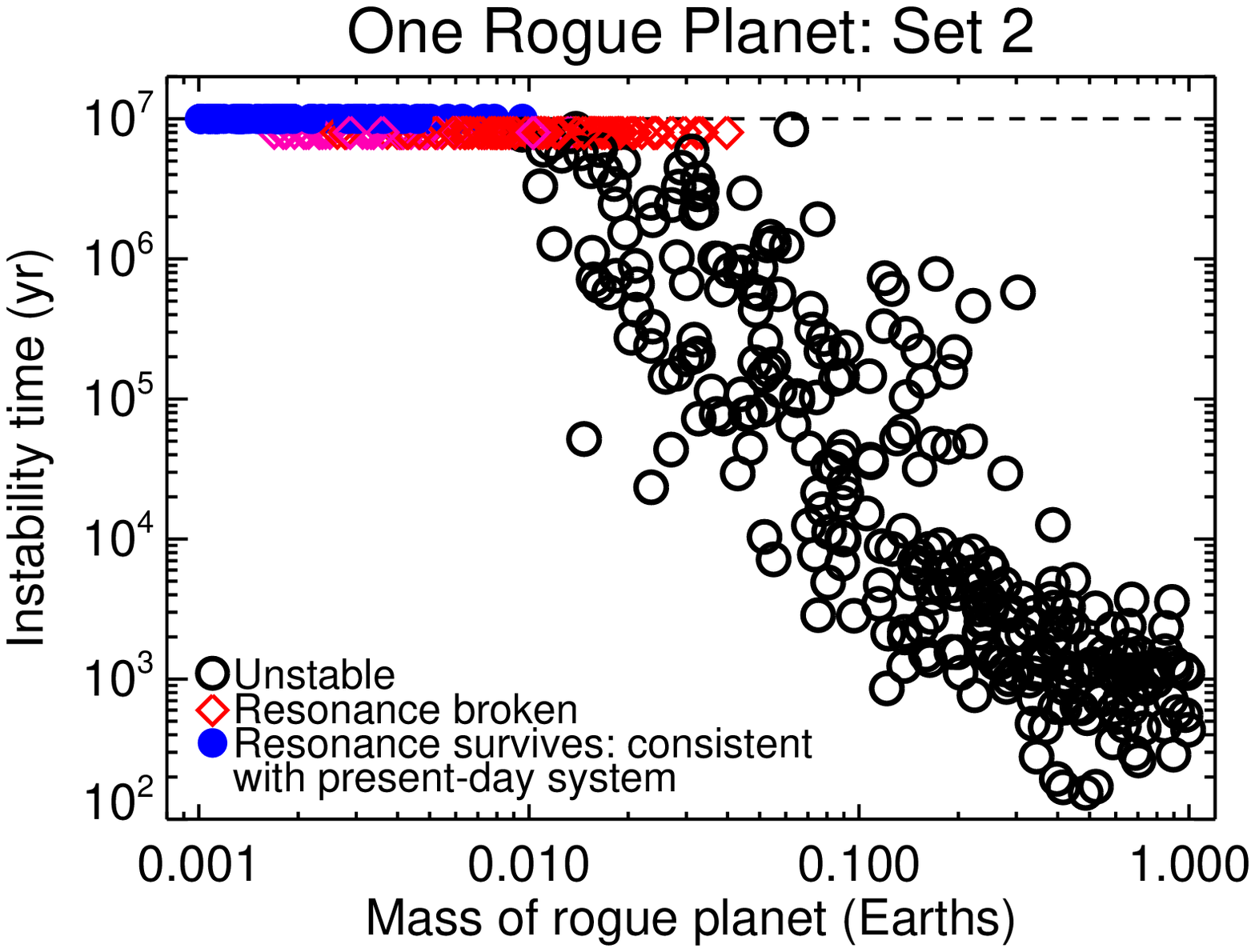} 
  \leavevmode \epsfxsize=6.0cm\epsfbox{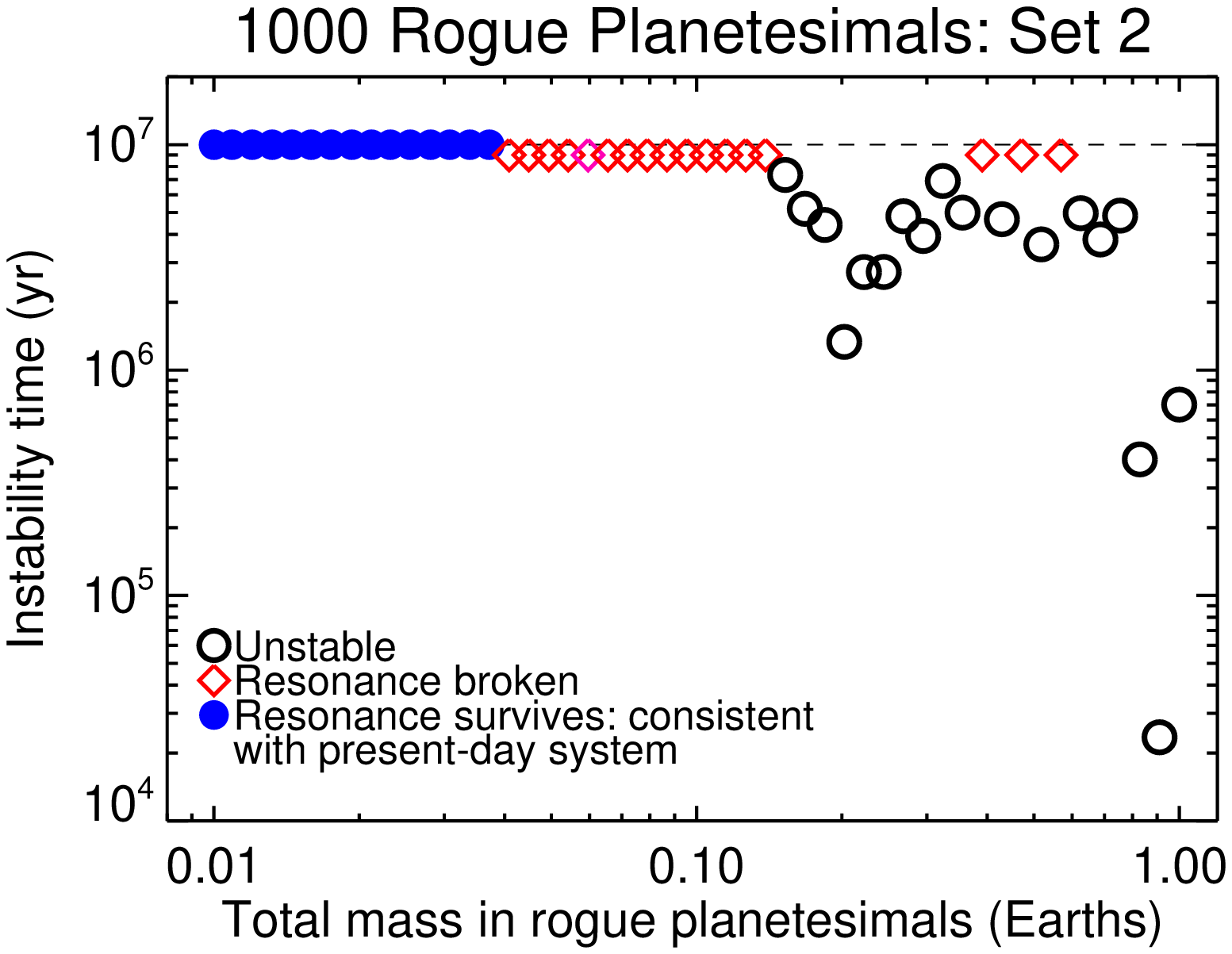}\\
      \leavevmode \epsfxsize=6.0cm\epsfbox{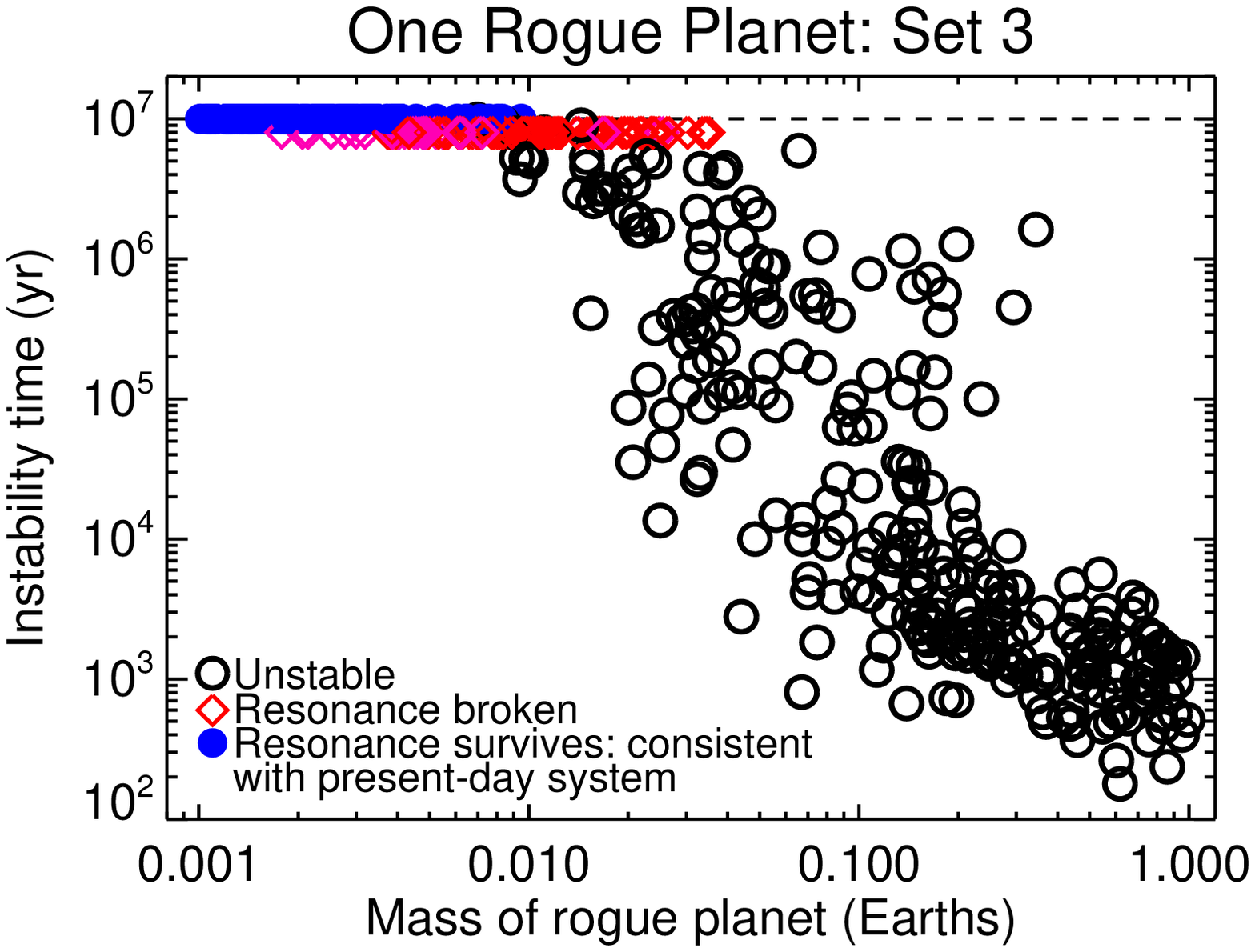} 
  \leavevmode \epsfxsize=6.0cm\epsfbox{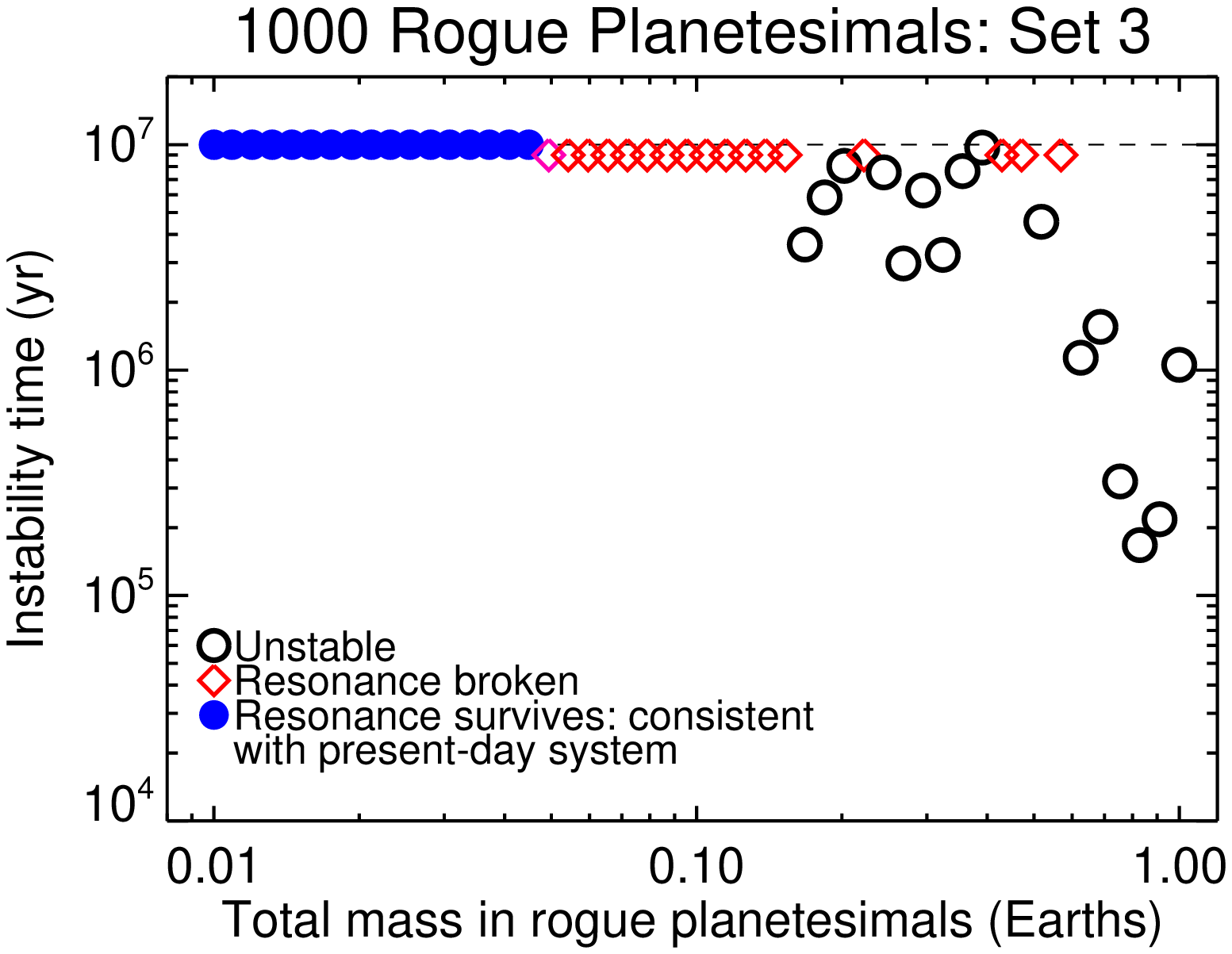}\\
      \leavevmode \epsfxsize=6.0cm\epsfbox{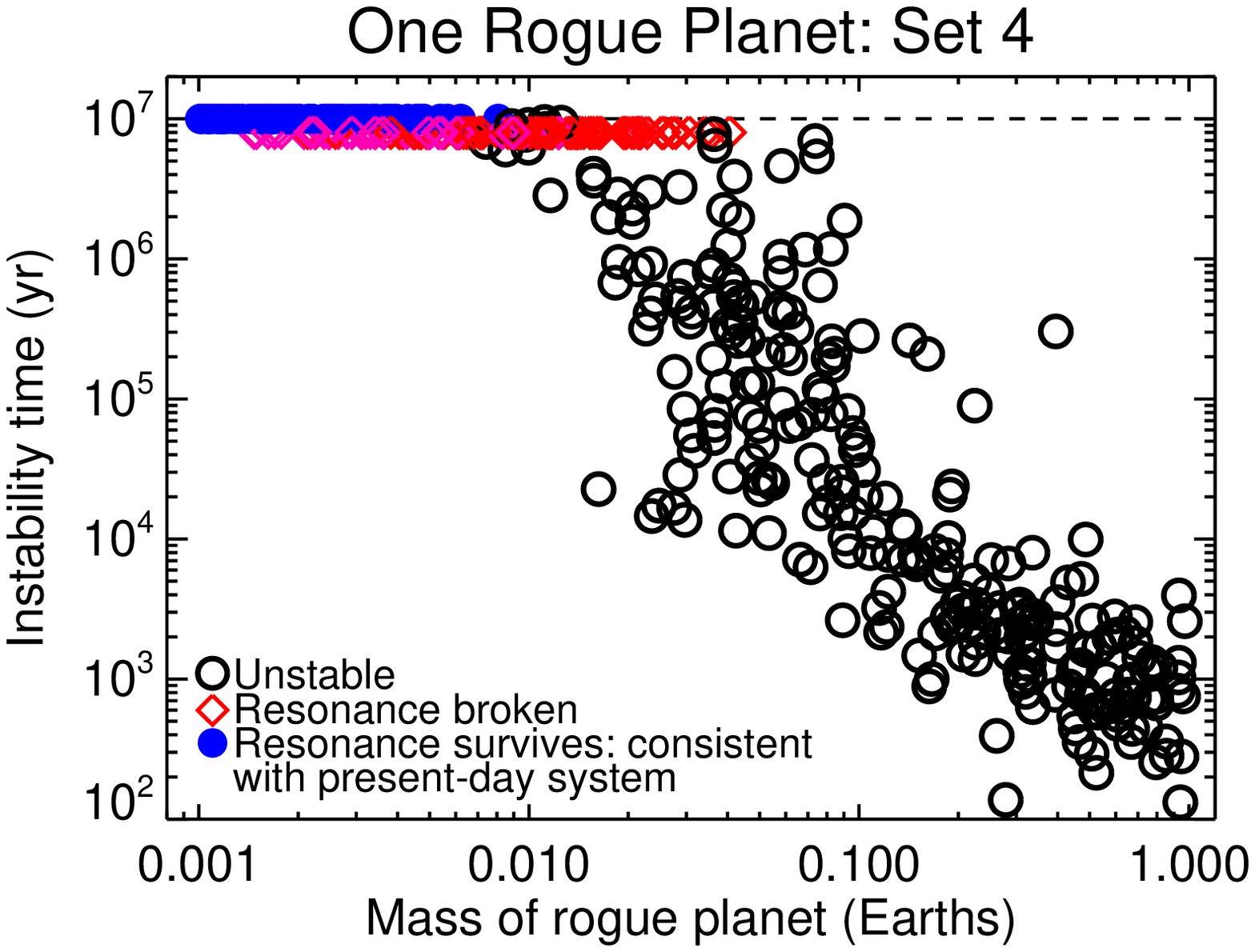} 
  \leavevmode \epsfxsize=6.0cm\epsfbox{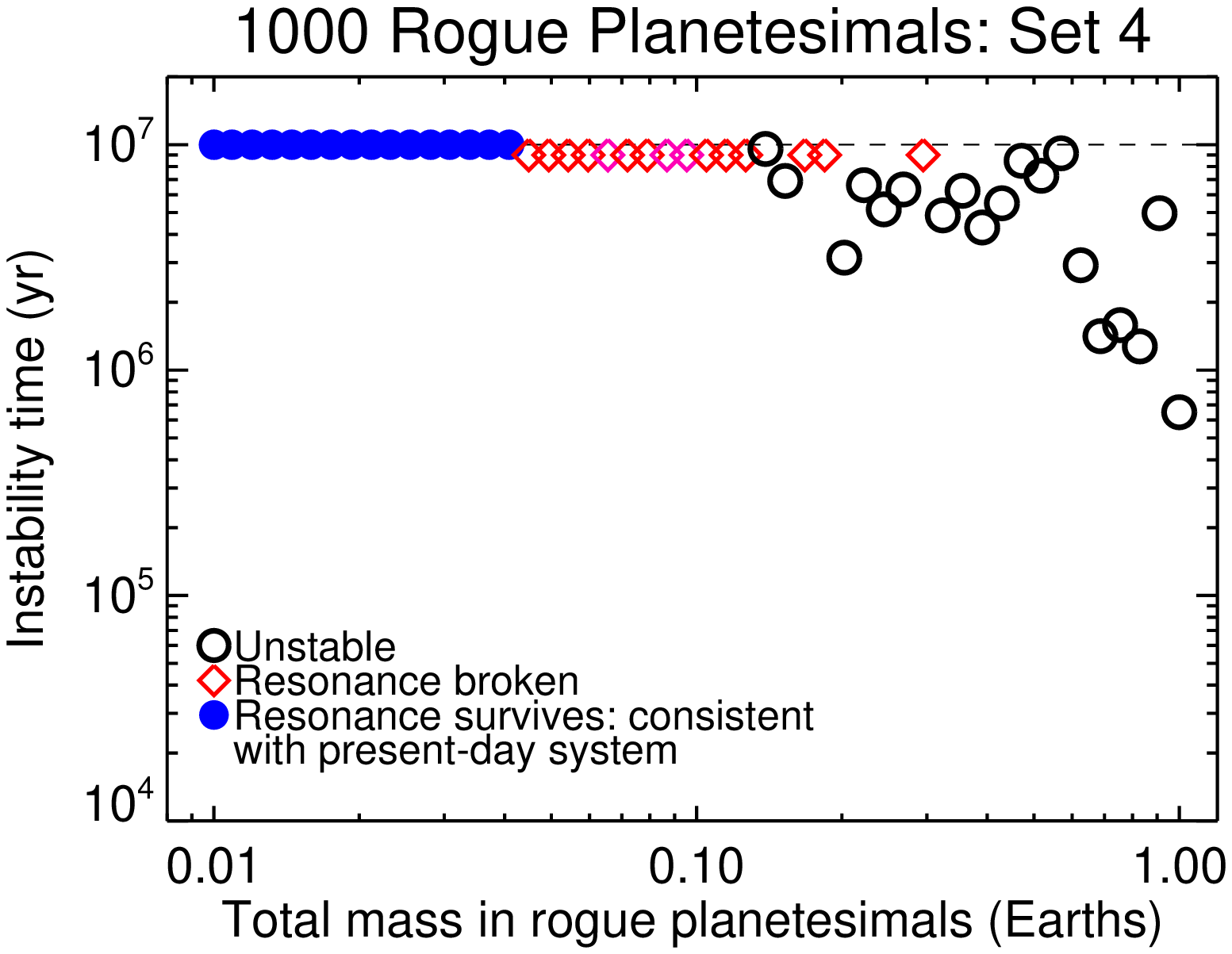}
    \caption*{{\bf Extended Data Fig. 5. }Effect of the assumed configuration of the Trappist-1 system. The left-hand panels show the outcomes of simulations containing a single rogue planet, for four different ``best-fit'' configurations of the Trappist-1 system that are consistent with current observations.\cite{agol21}  Likewise, the right-hand panels show the outcomes of simulations containing 1000 rogue planetesimals.  As in Fig.~1 in the main text, systems that were stable for 10 Myr but in which resonances were broken (red diamonds) are shifted downward slightly for clarity. The top panels (``Set 1'') represent our fiducial simulations (note that we have only used the subset of fiducial simulations with 1000 rogue planetesimals that fall in the identical mass range as the simulations for the other sets).  
    }
     \label{fig:inittest}
    \end{center}
\end{figure*}

Extended Data Fig.~5 shows the outcome of these sets of simulations, which are virtually identical to our fiducial set (referred to as Set 1).  The critical mass in planetesimals above which resonance was destroyed was always in the range $0.04-0.05 \mearth$. For the case of a single rogue planet, the results of our additional simulations are again very close to those for our fiducial set.  In fact, the results are slightly more restrictive in terms of the maximum mass of a late rogue planet: not a single simulation maintained its multi-resonant structure with a rogue planet more massive than $0.01 \mearth$. From all sets together, only a single simulation remained in a multi-resonant configuration after interacting with a rogue planet more massive than $0.01 \mearth$ and even in that case (in Set 1) the rogue planet was barely more massive than the Moon ($M_{Rogue} = 0.0167 \mearth = 1.4 M_{Moon}$).  That simulation had an unusual evolutionary path, as the rogue planet passed farther than 50~au from the star and was considered to be ejected from the system. The rogue planet was ejected in just $\sim 2\%$ of our simulations.  We therefore consider our upper limit for a single late giant impactor of a lunar mass to be robust.


\section*{6. Distribution of collision parameters}
We calculated the collisional parameters for each planetesimal collision that took place within simulations in which the planets remained in resonance. Almost all collisions took place within the first 100 kyr of each simulation, with a median collision timescale of $\sim 5-10$~kyr for each planet, consistent with previous work.\cite{kral18} Extended Data Fig.~6 shows the distribution of these impacts for our fiducial set of simulations (top panel, with initial orbital semimajor axes for rogue planetesimals of $a_{rogue} = 0.07-0.1$~au), as well as for the simulations with larger initial semimajor axes described above (see Methods Section 4 and Extended Data Fig.~4).  

\begin{figure*}
  \begin{center} 
  \leavevmode \epsfxsize=8cm\epsfbox{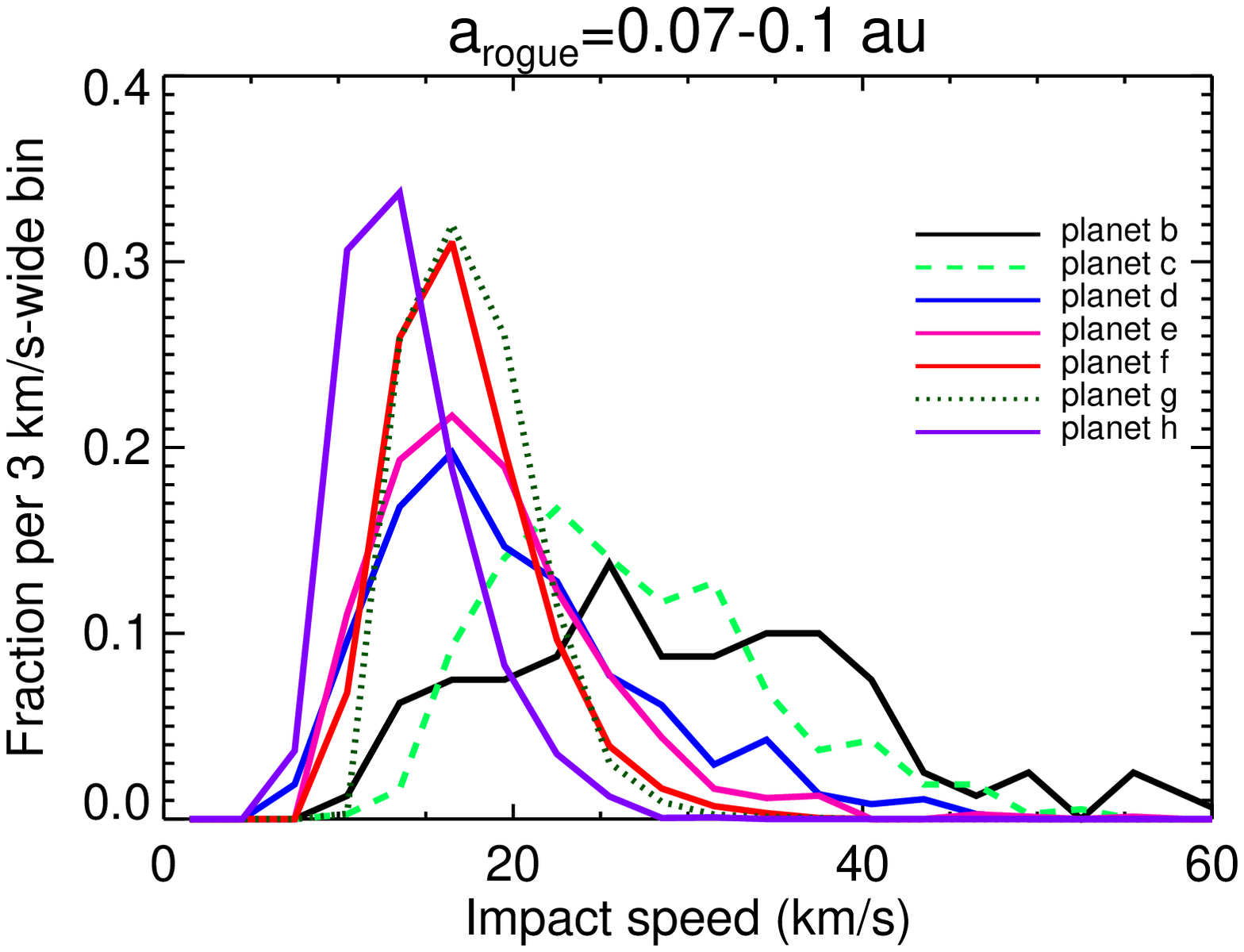}\\
  \leavevmode \epsfxsize=8cm\epsfbox{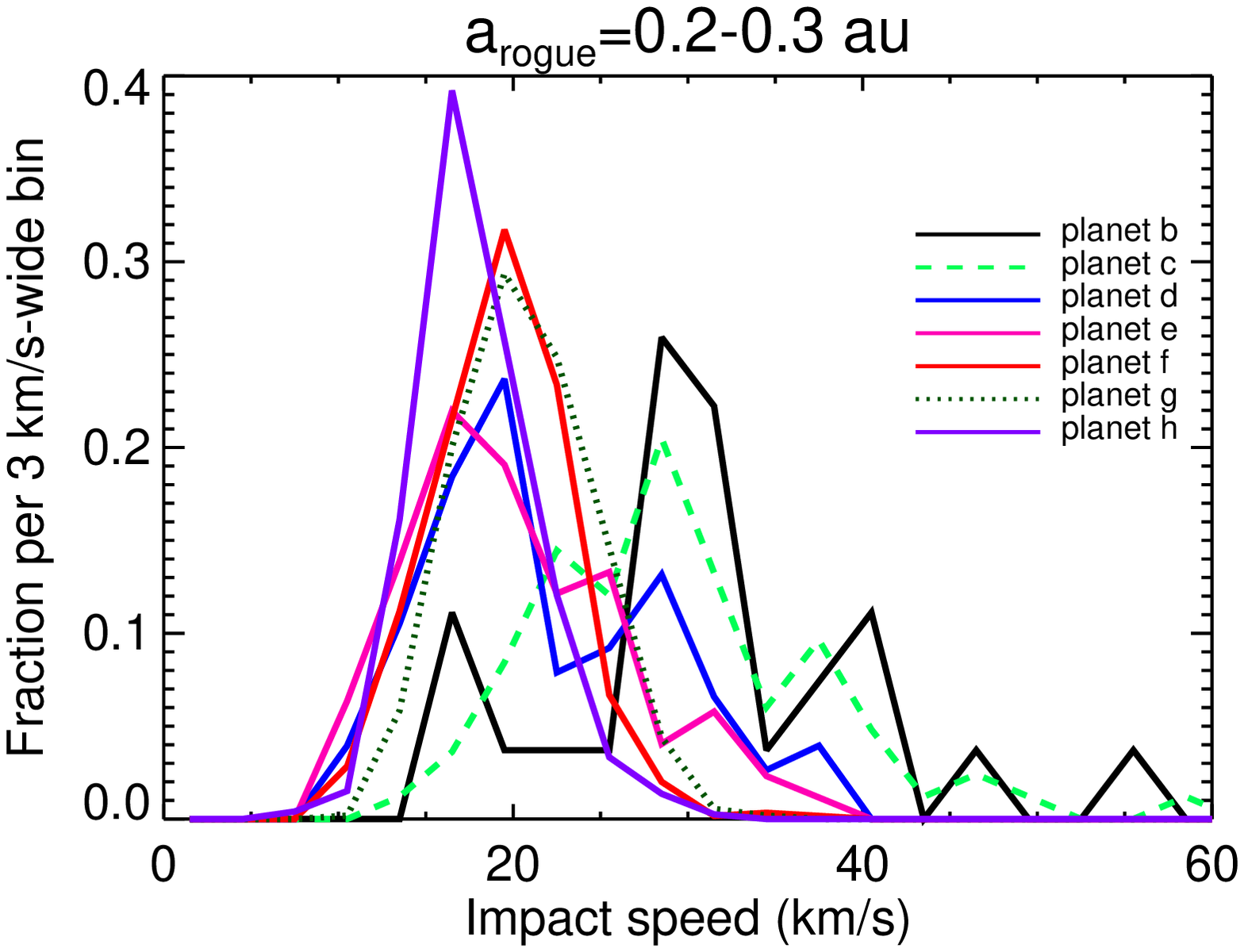}\\
  \leavevmode \epsfxsize=8cm\epsfbox{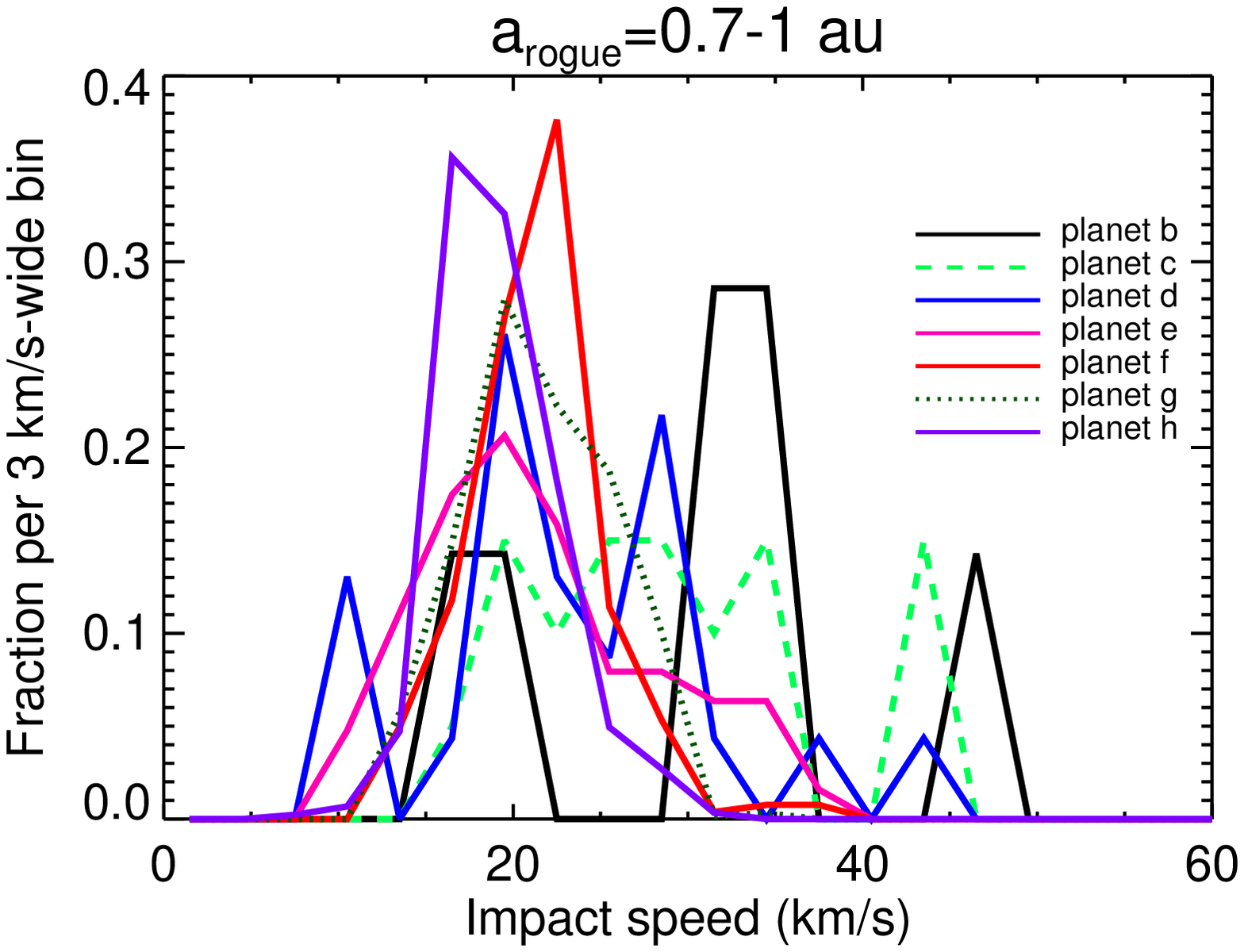}
    \caption*{{\bf Extended Data Fig. 6. }Distribution of impact speeds of planetesimals in simulations in which the planets remained in resonance on each of the Trappist-1 planets.  } 
     \label{fig:collparam}
    \end{center}
\end{figure*}

There are two trends in Extended Data Fig.~6.  The first is that the impact speed increases monotonically for planets closer to the star. Among just the fiducial set of simulations (top panel), median impact speeds onto the planets varied from 30 km/s for the innermost planet (b) to 13 km/s for the outermost (h), which is consistent with results of previous studies for similar impacting populations.\cite{kral18,dosovic20} The second trend is that impact speeds are modestly higher for larger initial orbital semimajor axis of the rogue objects.  For instance, the median impact velocities on planets g are 17.2, 20.3, and 21.1 km/s for increasing values of $a_{rogue}$ (from the top to bottom panels).

Both of these trends can be explained with simple analytical arguments. If we assume that impacts happen when a rogue object is at periastron, the rogue object's velocity is roughly the local Keplerian velocity $v_K = \sqrt{G M_\star/a}$ times $\sqrt{1+e}$, where $e$ is the rogue object's orbital eccentricity. The expected encounter speed between a rogue object at periastron and a planet on a roughly circular orbit at the same orbital radius is therefore $v_K (\sqrt{1+e}-1)$ (assuming parallel velocity vectors), and this value should represent a typical impact speed.\cite{kral18} This is indeed the case in our simulations. The scaling of $v_K$ with orbital radius can explain why, within a given system, impacts on the inner planets occur at higher speed that on the outer planets.  The trend of higher impact speeds for larger initial rogue semimajor axes $a_{rogue}$ can be explained by the second term, as rogue objects had fixed initial periastron distances such that larger $a_{rogue}$ corresponds to larger eccentricity.

Kral et al\cite{kral18} found a bimodal distribution of impact velocities on the Trappist-1 planets (see their Section 3.4).  Their lower-velocity impacts came from the same type of impacts seen in our simulations, in which collisions tended to take place when rogue objects were at periastron and their velocities were roughly parallel to the planets' orbits.  They also found a separate class of impacts from incoming high-eccentricity objects on roughly radial trajectories for which the velocity vectors were roughly perpendicular. In this case the expected impact speed should be $\sim v_K \sqrt{2+e}$.  We did not find any examples of such high-velocity collisions among the tens of thousands of impacts in our simulations; the origin of this discrepancy is almost certainly due to differing assumptions related to the impacting populations, specifically the initial periastron distributions of the impactors. Nonetheless, we did see a broader distribution in the impact speeds on the inner planets (see Extended Data Fig.~6).  This may be explained by the fact that the inner planets have higher collision probabilities with objects on a wide range of periastron distances whereas the outer planets are only likely to collide with rogue objects with periastron distances close to their orbital radius (see Fig.~2 in Kral et al\cite{kral18}).  The inner planets -- especially planets b and c -- are the most susceptible to impacts with non-parallel velocity vectors and therefore to a broader distribution of impact speeds.

\section*{7. Tidal evolution}

One might wonder whether resonances could be re-formed by orbital migration driven by tidal interactions between the planets and the central star. 
In order to obtain a convergent migration with tides, one needs an outward migration. 
The tides raised by planets g and h in the star can drive an outward migration due to the fact that the rotation period of Trappist-1 is smaller than the orbital period of the two planets, placing them outside the corotation radius.

To evaluate this possibility, we calculated separately the stellar tide induced migration for planet g and h using an equilibrium tide formalism\cite{2011A&A...535A..94B}, which accounts for the fact that the stellar radius evolves with time (we used the evolution tracks of Baraffe et al. (2015)\cite{2015A&A...577A..42B} for a stellar mass of $0.09~M_\odot$).
We computed as a function of time the evolution of the quantity $\Delta a$, which is the evolution of the separation between the two planets $a_h(t) - a_g(t)$ with respect to the initial separation $a_h(0) - a_g(0)$.
No tidal migration would lead to $\Delta a = 0$, a convergent migration would lead to $\Delta a < 0$ and a divergent migration to $\Delta a > 0$.
Due to the fact that the radius of a low mass star decreases significantly during the pre-main sequence, the strength of the stellar tide also decreases. 
Therefore depending on the initial time, we can expect different migration timescales.
Besides, the migration timescale highly depends on the assumed stellar dissipation.
Which is why we treat stellar dissipation as a free parameter and explore a range of $[10^{-3},...,10^3]$ times the dissipation of a fully convective star\cite{2010ApJ...723..285H,2012A&A...544A.124B}. 
Note that due to the lack of a radiative core for low-mass stars, we do not expect the dynamical tide to play a role as important as for Sun-like stars\cite{2016CeMDA.126..275B}.
Looking at a very wide range of dissipation is therefore a way to encompass the potential additional effect of inertial waves propagating in the star.

 \begin{figure*}
  \begin{center} 
   \leavevmode \epsfxsize=15cm\epsfbox{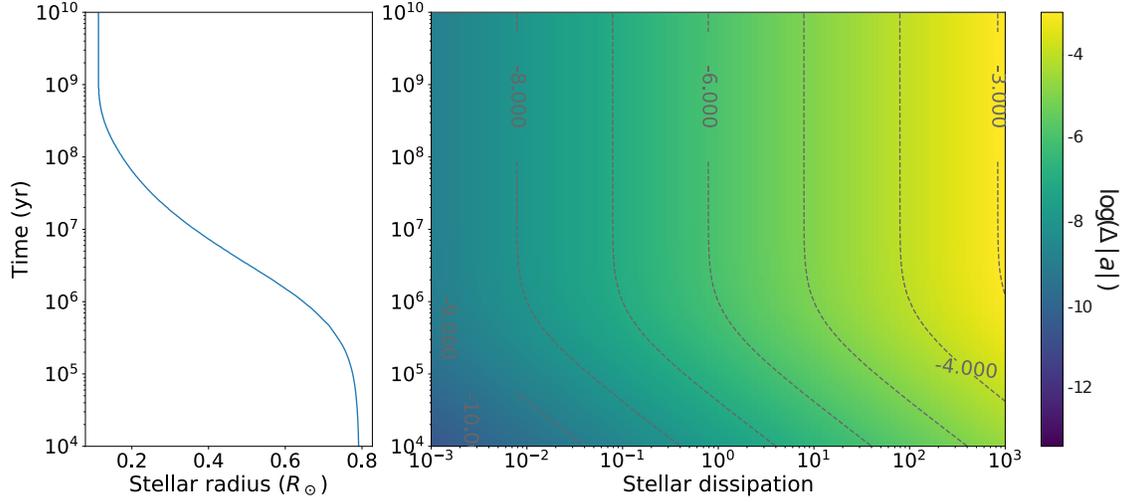}
    \caption*{{\bf Extended Data Fig. 7. } Effect of star-planet tidal interactions on the outer planets in the Trappist-1 system. Left panel: Evolution of the stellar radius with time (given as $t - t_{init}$, where $t_{init} = 3$~Myr). Right panel: map showing the values of $\log \Delta |a|$ as a function of the stellar dissipation and time. The stellar dissipation is given with respect to a fiducial value corresponding to a fully convective star or brown dwarf. 
        } 
     \label{fig:tides}
    \end{center}
\end{figure*}

The results for an initial time of $3$~Myr can been seen in Extended Data Fig.~7.
Due to the decrease of the stellar radius, most of the convergent migration occurs in the first million years. 
After that, the evolution timescales become too long and the evolution freezes.
Therefore, if the convergent migration was not enough to bring the planets back in resonance after a few million years, it never does.
We find that unless the star has a dissipation higher than 80 times the fiducial value, the convergent migration does not exceed $10^{-4}$~AU and is thus insufficient to bring the planets back in resonance.

Our reasoning assumes that the inner planets do not play a role on the outward migration of planet g and h, however in reality all planets will be submitted to the stellar tide-induced migration. Therefore, for completeness, we performed full N-body simulations including tides (using the Posidonius code\cite{2017ewas.confE...8B,2020A&A...635A.117B}) for different values of the stellar and planetary dissipation. We found that for stellar dissipations as low as the fiducial value, the system is destabilized. In particular, as planet b is closer than the corotation radius, it collides with the star before the star has time to spin down. This means that the dissipation needed to drive a sufficient convergent migration of planets g and h is not compatible with the system as we observe it today.

The star's spin rate -- and therefore the location of the corotation radius -- at early times is highly uncertain.\cite{herbst07}  If the star spun fast enough during its early evolution, the corotation radius would have been very close-in and tidally-induced migration could have been directed outward for all of the planets (see Bolmont et al\cite{bolmont12} for a discussion of the effect of the stellar spin history on tidal migration). While the calculations presented here do not account for that possibility, it is something to be addressed in future, detailed modeling of the tidal evolution of the system.

We have also assumed here that only star-planet tides influence the orbits of the planets, meaning that we have neglected other processes like planet-planet tides. However, it has been shown by Hay \& Matsuyama (2019)\cite{2019ApJ...875...22H} that the migration planet-planet tides induce is negligible compared to the star-planet tides. Moreover, planet-planet tides would induce an inward migration, which would rather lead to a spreading of the orbits rather than the convergent migration we are investigating, making the convergent migration of planets g and h even more challenging.

\section*{8. Internal structure modeling to estimate the planets' water contents}

We estimated the water mass fraction of each of the Trappist-1 planets as constrained by their measured sizes\cite{gillon16,gillon17} and masses.\cite{grimm18,agol21}  Each of the models (i-v) assumes uniform rocky interiors among all seven planets. This is because within a system, planets are expected to be similar in relative refractory abundances.\cite{dorn15,unterborn18,sotin}

The assumptions made in the five models were:
\begin{itemize}
    \item In model (i) we assumed a rocky interior with a core mass fraction of 23.5\% that best fits the measured densities of planets b, c, and d.
    \item In model (ii) we assumed rocky interiors with a core mass fraction of 26.6\% that best fit the measured density of the planet with the highest (uncompressed) density, planet c.
    \item In model (iii) we assumed bulk molar Fe/Mg of 0.75 and Mg/Si of 1.02 that was suggested by Unterborn et al\cite{unterborn18} to represent a proxy for the rocky interiors of the Trappist-1 planets.
    \item In model (iv) we constrained rocky interiors to solar ratios of Fe/Mg = 0.832 and Mg/Si = 0.886.\cite{lodders2009}
    \item In model (v), we use identical abundance constraints as in model (iii) but assume a core-free interior with all iron oxidised in the mantle. 
\end{itemize}

As suggested by Dorn et al.\cite{dorn18}, the densest planet may be a good proxy for the rocky components of other planets within the same system. This is the foundation of model (ii), for which the rocky interior is constrained by planet c's density (for a core mass fraction of 26.6 \%). Model (i) is an adaption of model (ii) but assumes that the densities of the three inner, least volatile-rich planets are representative of the interiors of all seven planets (for a core mass fraction 23.5 \%). Atmosphere models\cite{turbet20aa} indeed suggest that the three inner planets may be depleted in water and other volatiles\cite{agol21} (see also Fig.~3), which justifies using the density of these planets as a proxy for the interior composition of the outer planets. Models (iii) and (v) are both constrained by the stellar proxy\cite{unterborn18} that is based on refractory abundances of F-G-K stars with metallicities similar to Trappist-1. While model (iii) assumes all of each planet's iron is found in the core, model (v) has all the planets' iron oxidized in their mantles with core-free interiors, in line with recent estimates of oxygen budgets of extrasolar rocks.\cite{doyle2019oxygen} Model (iv) is for reference and assumes solar relative refractory abundances. 

The tested models (i-v) were then used to calculate the possible water mass fractions (wmf) needed to fit the planetary mass and radius data as well as stellar irradiation. To accomplish this we calculated the wmf for all planetary mass-radius pairs of the 10$^4$ posterior samples as in Agol et al.\cite{agol21}, thereby accounting for non-Gaussian distributed data errors. Whenever the sample's density is equal or higher than the pure rocky interior, we set the wmf of the sample to zero.

Interior profiles are calculated similar to previous studies.\cite{Dorn17,agol21} Specifically, we use equations of state for pure iron (hcp),\cite{hakim} mantle silicates,\cite{sotin} solid or liquid water\cite{haldemann} and steam water.\cite{Haar:1984} Following Agol et al.\cite{agol21}, water is treated as a condensed layer (liquid and/or solid) for planets less irradiated than the runaway greenhouse insolation threshold (i.e., Trappist-1 e,f,g,and h) ; water is treated as steam for planets more irradiated than the runaway greenhouse insolation threshold (i.e., Trappist-1 b,c,d),\cite{turbet19aa,turbet20aa} The atmospheric structure and radiative balance were computed using the LMD Generic 1-D climate model.\cite{turbet19aa,turbet20aa}


We neglected hydrated mantle rocks as they only affects the planet radius by less than 2 \% at maximum\cite{shah2020internal}, such that water is assumed to only be present in ice layers, oceans, and/or an atmosphere. Also, we do not show interiors with light core alloys, as our tested interiors with 70\% FeS in the core only lead marginal changes in planet radii ($<$1\%) compared with pure Fe cores while keeping the same bulk Fe/Si ratios.

Tables containing all of the results of our modeling are included in the Supplementary Information (Supplementary Tables 2-6).

\section*{9. Description of the simulation from Fig.~2 in the main text that formed our illustrative Trappist-1 `analog.'}

The Trappist-1 analog planets in Figure 2 from the main text represent a super-Earth system produced in simulations of Izidoro et al\cite{izidoro21}. Their planet formation model includes the effects of pebble accretion, type-I migration, eccentricity and inclination damping due to tidal interaction of protoplanetary objects with the disk.  This particular system shown was produced in the setup of Model-III of Izidoro et al. It started with $\sim 70$ Moon-mass planetary seeds in the gas disk, which were initially placed on orbital radii between 0.2~au and 2~au and radially separated from each other by 0.025~au. The inner edge of the gas disk was located at 0.1~au. At the beginning of the simulation, the  snow line was beyond 2~au. Seeds first grew via pebble accretion (interior to the snow line), then type-I migrated, and eventually underwent mutual collisions.  As the disk evolved the snowline moved inwards, eventually sweeping interior to the orbits of the outermost seeds and allowing them to accrete water-ice rich pebbles. The outermost seeds grew fast enough to regulate the pebble flux to the inner ones. During the 5~Myr gaseous disk lifetime a system of 7 planets formed in a multi-resonant configuration between 0.05~au and 0.24~au.  The planets' masses were 0.77~$M_{\oplus}$, 0.92~$M_{\oplus}$,  1.03~$M_{\oplus}$, 1.42~$M_{\oplus}$, 1.09~$M_{\oplus}$, 0.49~$M_{\oplus}$, and 0.42~$M_{\oplus}$. The period ratio of adjacent planet pairs were approximately, 1.25, 1.33, 1.33, 1.33, 1.25, and 2.42 (the two outer planets show libration of certain angles associated with the 12:5 and 5:2 resonances). The four innermost planets formed dry, meaning that they did not accrete water-rich pebbles or collide with any water-carrying seeds during the entire course of the simulation. The fifth planet contained $\sim 14\%$ water by mass, the sixth planet was dry, and the outermost planet contained $\sim 20\%$ water by mass. The disk dissipated after 5~Myr and we continued the simulation in  a gas-free environment for additional 50~Myr to verify that the resonant configuration remained dynamically stable.

The primary goal of Figure 2 in the main text is to illustrate one possible pathway for the rapid growth of Trappist-1 like systems. This particular system is clearly not a perfect match to the Trappist-1 system. First of all, the resonant structure is not quite right, as the system contains too many 5:4 and 4:3 mean motion resonances, rather than 3:2 (see Papaloizou et al\cite{papaloizou18}, Coleman et al\cite{coleman19} and Lin et al\cite{lin21} for careful studies of the resonant structure of simulated Trappist-1 analogs).  In addition, the two planets with $>10\%$ water by mass are at the (upper) edge of the allowed parameter space in the interior structure models presented in Fig.~3.  Nonetheless, our chosen Trappist-1 analog does have outstanding similarities to the Trappist-1 system such as the high planet multiplicity and resonant state of adjacent planets.  The simulation also shows that planet formation models can qualitatively account for the different bulk compositions of the Trappist-1 planets as predicted by interior models.

\vskip .2in
\noindent{\bf Code availability}\\
The {\tt Mercury} N-body integrator\cite{chambers99} is publicly-available and can be downloaded at \url{https://github.com/4xxi/mercury}.  We made trivial modifications to stop simulations in which two planets collided and to create an additional data dump any time a planetesimal was removed from the simulation.  For water content calculations, the interior structure code is available from C.D. (cdorn@physik.uzh.ch) upon reasonable request, and the atmospheric code is based on polynomial fits from Turbet et al (2020).\cite{turbet20aa} 

\noindent{\bf Data availability}\\
All simulation data and model data that support the findings of this study or were used to make the plots are available from the corresponding author upon reasonable request.

\end{methods}

\begin{addendum}
 \item [Dedication] This paper is dedicated to the memory of our friend and colleague Franck Hersant. 
 \item [Acknowledgments] We thank Sylvie Rousseau and Arnaud Collioud for managing the machines on which our simulations were run. S.~N.~R. is grateful to the ECLIPSE research team at the Laboratoire d'Astrophysique de Bordeaux and the ``Zooming In On Rocky Planet Formation'' ISSI Team for helpful discussions, and to Aurelie Guilbert-Lepoutre and Anastasis Gkotsinas for input on the water contents of comets. S.~N.~R. and F.~S. thank the CNRS's PNP program for support. A.~I and R.~D. acknowledge NASA grant 80NSSC18K0828 for financial support during preparation and submission of this work. A.~I. also thanks the Brazilian Federal Agency for Support and Evaluation of Graduate Education (CAPES), in the scope of the Program CAPES-PrInt, process number 88887.310463/2018-00, International Cooperation Project number 3266, and CNPq (313998/2018-3). P.~B. acknowledges a St Leonard's Interdisciplinary Doctoral Scholarship from the University of St Andrews. L.C. acknowledges support from the DFG Priority Programme SP1833 Grant CA 1795/3. E.A. and S.N.R. acknowledge support from from the Virtual Planetary Laboratory Team, a member of the NASA Nexus for Exoplanet System Science, funded via NASA Astrobiology Program Grant No. 80NSSC18K0829. This project has received funding from the European Union’s Horizon 2020 research and innovation program under the Marie Sklodowska-Curie Grant Agreement No. 832738/ESCAPE. This work has been carried out within the framework of the National Centre of Competence in Research PlanetS supported by the Swiss National Science Foundation. M.T. and E.B. acknowledge the financial support of the SNSF. This work was performed using the High-Performance Computing (HPC) resources of Centre Informatique National de l'Enseignement Supérieur (CINES) under the allocation A0080110391 made by Grand Équipement National de Calcul Intensif (GENCI). M.T. thanks the Gruber Foundation for its support to this research. C.D. acknowledges support from the Swiss National Science Foundation under grant PZ00P2\_174028.
 \item [Author Contributions] S.~N.~R. initiated the project after discussions with L.~C., P.~B. and A.~I., ran and analyzed the simulations using codes optimized by A.~I. and configurations of the system provided by S.~G. and E.~A..  C.~D. and M~T. performed the interior modeling to obtain constraints on the water contents of the Trappist-1 planets. E.~B. performed the analysis of tidal effects. S.~N.~R. wrote the paper with feedback from A.~I., E.~B., C.~D., F.~S., M.~T., E.~A., P.~B., L.~C., R.~D., M.~G., and S.~G. 
\item [Competing Interests] The authors declare no competing interests.
 \item[Correspondence] Correspondence and requests for materials should be addressed to S.N.R (rayray.sean@gmail.com).
\end{addendum}

\newpage

\clearpage
\newpage

\clearpage
\newpage

\section*{\Large Supplementary Information}

\section*{Exact orbital configurations of the Trappist-1 system that were used in our analysis}

As explained in the Methods, we tested four different best-fit orbital configurations of the system, each drawn from Agol et al.\cite{agol21}  They are listed in Table 1.

\begin{table*}
\centering
\begin{tabular}{ c | c | c | c | c | c | c}
Planet & Mass  & Radius  & Semimajor & Eccentricity & Longitude of & Mean \\
 &  ($\mearth$) & ($\rearth$)& Axis $a$ (AU) & $e$ & periastron $\varpi$ ($^\circ$) & Anomaly $M$ ($^\circ$) \\
  \hline
  {\bf Fiducial (Set 1)}\\
  b & 1.3925 & 1.1174 & 0.011551 & 0.002344 & 253.61247 & 105.78489 \\
  c & 1.2943 & 1.0967 & 0.015820 &  0.001224 & 132.62793 & 54.89836 \\
  d & 0.3958 & 0.7880 & 0.02229 & 0.005045 & 202.45580 & 171.39157 \\
  e & 0.6824 & 0.9200 & 0.02930 & 0.007013 & 52.42997 & 30.97582 \\
  f & 1.0634 & 1.0448 & 0.038551 & 0.008298 & 170.04247 & 247.44087 \\
  g & 1.3464 & 1.1294 & 0.046896 & 0.003760 & 355.97714 & 87.27858 \\
  h & 0.3198 & 0.7552 & 0.061963 & 0.003571 & 172.18673 & 118.58431  \\
  \hline  
  \hline
  {\bf Set 2}\\
  b & 1.3925 & 1.1174 & 0.011551 & 0.004677 & 181.37652 & 178.45848 \\
  c & 1.2896 & 1.0979 & 0.015821 &  0.002755 & 276.77544 & 270.77957 \\
  d & 0.3909 & 0.7889 & 0.022289 & 0.007069 & 217.02820 & 156.90568 \\
  e & 0.7068 & 0.9210 & 0.029296 & 0.005359 & 44.52192 & 38.91996 \\
  f & 1.0477 & 1.0460 & 0.038551 & 0.009050 & 179.50878 & 238.00290 \\
  g & 1.3232 & 1.1307 & 0.046896 & 0.003686 & 335.57244 & 107.74799 \\
  h & 0.3309 & 0.7561 & 0.061965 & 0.003714 & 188.97120 & 101.87152  \\
  \hline  
  \hline	
    {\bf Set 3}\\
  b & 1.3405 & 1.1174 & 0.011550 & 0.003402 & 194.66082 & 165.01640 \\
  c & 1.3062 & 1.0979 & 0.015818 &  0.000802 & 196.57200 & 351.02606 \\
  d & 0.3987 & 0.7889 & 0.022290 & 0.006354 & 210.05726 & 163.88358 \\
  e & 0.7003 & 0.9210 & 0.029295 & 0.005947 & 51.87353 & 31.58265 \\
  f & 1.0777 & 1.0460 & 0.038549 & 0.008950 & 175.24378 & 242.26801 \\
  g & 1.3652 & 1.1307 & 0.046897 & 0.003526 & 347.44533 & 95.88219 \\
  h & 0.3404 & 0.7561 & 0.061966 & 0.003855 & 172.27644 & 118.53050  \\
  \hline  
  \hline  
    {\bf Set 4}\\
  b & 1.3370 & 1.1174 & 0.011551 & 0.000611 & 167.48106 & 191.83734 \\
  c & 1.2811 & 1.0979 & 0.015818 &  0.004240 & 330.78097 & 216.35048 \\
  d & 0.3803 & 0.7889 & 0.022288 & 0.005124 & 210.85020 & 162.86095 \\
  e & 0.7112 & 0.9210 & 0.029294 & 0.007072 & 45.29093 & 38.04304 \\
  f & 1.0464 & 1.0460 & 0.038551 & 0.007802 & 173.00821 & 244.39623 \\
  g & 1.3231 & 1.1307 & 0.046892 & 0.004384 & 352.07474 & 91.11644 \\
  h & 0.3050 & 0.7561 & 0.061960 & 0.003833 & 182.31057 & 108.43892  \\
  \hline  
\end{tabular}
\caption{Starting orbital configurations of the Trappist-1 system that were used in our simulations. The fiducial system is at the top (and referred to as Set 1 in our tests).  In all systems the planets' orbits have zero inclinations, but in our simulations rogue planets and planetesimals were given non-zero inclinations to ensure that the systems were not 2-dimensional. Each system was taken from the best-fit posteriors of Agol et al.\cite{agol21} }
\end{table*}

\section*{Interior structure modeling to estimate the water contents of the Trappist-1 planets}

In Tables 2-6 we include the full results of our interior structure modeling described in the Methods.  Each of the five tables contains the derived water contents (including the means, 1-$\sigma$ and $2-\sigma$ errors) for each planet.

\begin{table*}
\centering
\begin{tabular}{ c | c | c | c }
Planet & Mean water content  & Uncertainties ($1\sigma$)  & Uncertainties ($2\sigma$) \\
 &  by mass &  &   \\
  \hline			
  b &  -- & -- & -- \\
  c & -- & -- & -- \\
  d &  -- & -- & -- \\
  e & $6.88\times 10^{-3}$ & $^{+1.58\times 10^{-2}}_{-6.88\times 10^{-3}}$ & 
  $^{+4.31\times 10^{-2}} _{-6.88\times 10^{-3}}$\\
  f &  $2.05\times 10^{-2}$  & $^{+1.69\times 10^{-2}}_{-1.20\times 10^{-2}}$ & $^{+4.40\times 10^{-2}}_{-2.05\times 10^{-2}}$ \\
  g & $3.64\times 10^{-2}$  & $^{+1.84\times 10^{-2}}_{-1.42\times 10^{-2}}$   &   $^{+4.82\times 10^{-2}} _{- 2.66\times 10^{-2}}$\\
  h & $3.53\times 10^{-2}$  & $^{+ 4.37\times 10^{-2}} _{-3.02\times 10^{-2}}$   &  $^{+9.57\times 10^{-2}} _{- 3.53\times 10^{-2}}$\\
  \hline  
\end{tabular}
\label{tab:water1}
\caption{Water contents from interior model i.}
\end{table*}

\begin{table*}
\centering
\begin{tabular}{ c | c | c | c }
Planet & Mean water content  & Uncertainties ($1\sigma$)  & Uncertainties ($2\sigma$) \\
 &  by mass &  &   \\
  \hline			
  b &  -- & -- & $^{+3.11\times 10^{-5}}$ \\
  c & -- & -- & -- \\
  d &  -- & -- & -- \\
  e & $1.60\times 10^{-2}$ &  $^{+1.88\times 10^{-2}}_{- 1.21\times 10^{-2}}$   &   $^{+4.66\times 10^{-2}}_{-1.60\times 10^{-2}}$ \\
  f & $3.25\times 10^{-2}$ & $^{+ 1.74\times 10^{-2}}_{- 1.36\times 10^{-2}}$   &   $^{+ 4.50\times 10^{-2}}_{-2.43\times 10^{-2}}$\\
  g & $4.90\times 10^{-2}$ & $^{+1.89\times 10^{-2}} _{-1.47\times 10^{-2}}$   &   $^{+ 4.94\times 10^{-2}} _{- 2.78\times 10^{-2}}$\\
  h & $4.70\times 10^{-2}$ &  $^{+4.46\times 10^{-2}} _{- 3.37\times 10^{-2}}$   &  $^{+9.65\times 10^{-2}} _{- 4.70\times 10^{-2}}$\\
  \hline  
\end{tabular}
\label{tab:water2}
\caption{Water contents from interior model ii.}
\end{table*}

\begin{table*}
\centering
\begin{tabular}{ c | c | c | c }
Planet & Mean water content  & Uncertainties ($1\sigma$)  & Uncertainties ($2\sigma$) \\
 &  by mass &  &   \\
  \hline			
  b &  -- & $^{+1.05\times 10^{-5}}$ & $^{+6.89\times 10^{-5}}$ \\
  c & -- & -- & $^{+2.69\times 10^{-5}}$ \\
  d &  -- & -- & -- \\
  e & $2.75\times 10^{-2}$ &  $^{+ 2.06\times 10^{-2}}_{-1.57\times 10^{-2}}$ &   $^{+ 5.03\times 10^{-2}}_{- 2.59\times 10^{-2}}$\\
  f & $4.59\times 10^{-2}$ & $^{+ 1.88\times 10^{-2}}_{  - 1.46\times 10^{-2}}$   &   $^{+ 4.84\times 10^{-2}}_{  - 2.76\times 10^{-2}}$\\
  g & $6.40\times 10^{-2}$ & $^{+ 2.03\times 10^{-2} }_{ - 1.60\times 10^{-2} }$  &   $^{+ 5.31\times 10^{-2}}_{  - 3.03\times 10^{-2}}$\\
  h & $6.11\times 10^{-2}$ & $^{+ 4.68\times 10^{-2} }_{ - 3.79\times 10^{-2} }$  &   $^{+ 1.01\times 10^{-1}}_{  - 6.11\times 10^{-2}}$\\
  \hline  
\end{tabular}
\label{tab:water3}
\caption{Water contents from interior model iii.}
\end{table*}

\begin{table*}
\centering
\begin{tabular}{ c | c | c | c }
Planet & Mean water content  & Uncertainties ($1\sigma$)  & Uncertainties ($2\sigma$) \\
 &  by mass &  &   \\
  \hline			
  b &  -- & $^{+1.65\times 10^{-5} }$ & $^{+1.25\times 10^{-4} }$  \\
  c & -- & -- & $^{+5.11\times 10^{-5}  }$\\
  d &  -- & -- & -- \\
  e & $3.69\times 10^{-2} $& $^{+ 2.14\times 10^{-2} }_{ - 1.76\times 10^{-2} }$  &   $^{+ 5.17\times 10^{-2} }_{ - 2.96\times 10^{-2}}$ \\
  f & $5.61\times 10^{-2}$ & $^{+ 1.95\times 10^{-2} }_{ - 1.54\times 10^{-2} }$  &   $^{+ 5.00\times 10^{-2} }_{ - 2.87\times 10^{-2}}$ \\
  g & $7.49\times 10^{-2}$ & $^{+ 2.12\times 10^{-2} }_{ - 1.64\times 10^{-2} }$  &   $^{+ 5.56\times 10^{-2} }_{ - 3.14\times 10^{-2}}$\\
  h & $7.16\times 10^{-2}$ & $^{+ 4.77\times 10^{-2} }_{ - 4.02\times 10^{-2} }$  &   $^{+ 1.03\times 10^{-1 }}_{ - 6.91\times 10^{-2}}$\\
  \hline  
\end{tabular}
\label{tab:water4}
\caption{Water contents from interior model iv.}
\end{table*}

\begin{table*}
\centering
\begin{tabular}{ c | c | c | c }
Planet & Mean water content  & Uncertainties ($1\sigma$)  & Uncertainties ($2\sigma$) \\
 &  by mass &  &   \\
  \hline			
  b &  -- & -- &  \\
  c & -- & -- & -- \\
  d &  -- & -- & -- \\
  e & -- & $^{+2.55\times 10^{-2} }$  &  $^{+4.92\times 10^{-2} }$\\
  f & $7.01\times 10^{-3}$ &  $^{+1.39\times 10^{-2}}_{  - 7.01\times 10^{-3}}$   &   $^{+ 3.91\times 10^{-2} }_{ - 7.01\times 10^{-3}}$\\
  g & $2.01\times 10^{-2}$ & $^{+1.70\times 10^{-2} }_{ - 1.20\times 10^{-2} }$  &   $^{+ 4.50\times 10^{-2} }_{ - 2.01\times 10^{-2}}$\\
  h & $2.13\times 10^{-2}$ &  $^{+3.98\times 10^{-2} }_{ - 2.13\times 10^{-2}}$   &   $^{+ 9.04\times 10^{-2} }_{ - 2.13\times 10^{-2}}$\\
  \hline  
\end{tabular}
\label{tab:water5}
\caption{Water contents from interior model v.}
\end{table*}


\end{document}